\documentclass{aastex63}
\usepackage[T5,T1]{fontenc}

\usepackage{multirow}
\usepackage{amsmath}
\usepackage{amssymb}
\usepackage{mathtools}
\usepackage{tabularx}
\usepackage{longtable}
\usepackage{booktabs}
\usepackage{changepage}
\usepackage{lineno}

\newcommand{\MA}[1]{{\color{black}#1}}
\newcommand{\vect}[1]{\boldsymbol{#1}}

\shorttitle{Search for time-dependent neutrino emission from a source catalog}
\shortauthors{Aartsen et al.}
\graphicspath{{./}{figures/}}

\begin{document}

\title{Search for multi-flare neutrino emissions in 10 years of IceCube data from a catalog of sources}

\affiliation{III. Physikalisches Institut, RWTH Aachen University, D-52056 Aachen, Germany}
\affiliation{Department of Physics, University of Adelaide, Adelaide, 5005, Australia}
\affiliation{Dept. of Physics and Astronomy, University of Alaska Anchorage, 3211 Providence Dr., Anchorage, AK 99508, USA}
\affiliation{Dept. of Physics, University of Texas at Arlington, 502 Yates St., Science Hall Rm 108, Box 19059, Arlington, TX 76019, USA}
\affiliation{CTSPS, Clark-Atlanta University, Atlanta, GA 30314, USA}
\affiliation{School of Physics and Center for Relativistic Astrophysics, Georgia Institute of Technology, Atlanta, GA 30332, USA}
\affiliation{Dept. of Physics, Southern University, Baton Rouge, LA 70813, USA}
\affiliation{Dept. of Physics, University of California, Berkeley, CA 94720, USA}
\affiliation{Lawrence Berkeley National Laboratory, Berkeley, CA 94720, USA}
\affiliation{Institut f{\"u}r Physik, Humboldt-Universit{\"a}t zu Berlin, D-12489 Berlin, Germany}
\affiliation{Fakult{\"a}t f{\"u}r Physik {\&} Astronomie, Ruhr-Universit{\"a}t Bochum, D-44780 Bochum, Germany}
\affiliation{Universit{\'e} Libre de Bruxelles, Science Faculty CP230, B-1050 Brussels, Belgium}
\affiliation{Vrije Universiteit Brussel (VUB), Dienst ELEM, B-1050 Brussels, Belgium}
\affiliation{Department of Physics and Laboratory for Particle Physics and Cosmology, Harvard University, Cambridge, MA 02138, USA}
\affiliation{Dept. of Physics, Massachusetts Institute of Technology, Cambridge, MA 02139, USA}
\affiliation{Dept. of Physics and Institute for Global Prominent Research, Chiba University, Chiba 263-8522, Japan}
\affiliation{Department of Physics, Loyola University Chicago, Chicago, IL 60660, USA}
\affiliation{Dept. of Physics and Astronomy, University of Canterbury, Private Bag 4800, Christchurch, New Zealand}
\affiliation{Dept. of Physics, University of Maryland, College Park, MD 20742, USA}
\affiliation{Dept. of Astronomy, Ohio State University, Columbus, OH 43210, USA}
\affiliation{Dept. of Physics and Center for Cosmology and Astro-Particle Physics, Ohio State University, Columbus, OH 43210, USA}
\affiliation{Niels Bohr Institute, University of Copenhagen, DK-2100 Copenhagen, Denmark}
\affiliation{Dept. of Physics, TU Dortmund University, D-44221 Dortmund, Germany}
\affiliation{Dept. of Physics and Astronomy, Michigan State University, East Lansing, MI 48824, USA}
\affiliation{Dept. of Physics, University of Alberta, Edmonton, Alberta, Canada T6G 2E1}
\affiliation{Erlangen Centre for Astroparticle Physics, Friedrich-Alexander-Universit{\"a}t Erlangen-N{\"u}rnberg, D-91058 Erlangen, Germany}
\affiliation{Physik-department, Technische Universit{\"a}t M{\"u}nchen, D-85748 Garching, Germany}
\affiliation{D{\'e}partement de physique nucl{\'e}aire et corpusculaire, Universit{\'e} de Gen{\`e}ve, CH-1211 Gen{\`e}ve, Switzerland}
\affiliation{Dept. of Physics and Astronomy, University of Gent, B-9000 Gent, Belgium}
\affiliation{Dept. of Physics and Astronomy, University of California, Irvine, CA 92697, USA}
\affiliation{Karlsruhe Institute of Technology, Institute for Astroparticle Physics, D-76021 Karlsruhe, Germany }
\affiliation{Karlsruhe Institute of Technology, Institute of Experimental Particle Physics, D-76021 Karlsruhe, Germany }
\affiliation{Dept. of Physics, Engineering Physics, and Astronomy, Queen's University, Kingston, ON K7L 3N6, Canada}
\affiliation{Dept. of Physics and Astronomy, University of Kansas, Lawrence, KS 66045, USA}
\affiliation{Department of Physics and Astronomy, UCLA, Los Angeles, CA 90095, USA}
\affiliation{Department of Physics, Mercer University, Macon, GA 31207-0001, USA}
\affiliation{Dept. of Astronomy, University of Wisconsin{\textendash}Madison, Madison, WI 53706, USA}
\affiliation{Dept. of Physics and Wisconsin IceCube Particle Astrophysics Center, University of Wisconsin{\textendash}Madison, Madison, WI 53706, USA}
\affiliation{Institute of Physics, University of Mainz, Staudinger Weg 7, D-55099 Mainz, Germany}
\affiliation{Department of Physics, Marquette University, Milwaukee, WI, 53201, USA}
\affiliation{Institut f{\"u}r Kernphysik, Westf{\"a}lische Wilhelms-Universit{\"a}t M{\"u}nster, D-48149 M{\"u}nster, Germany}
\affiliation{Bartol Research Institute and Dept. of Physics and Astronomy, University of Delaware, Newark, DE 19716, USA}
\affiliation{Dept. of Physics, Yale University, New Haven, CT 06520, USA}
\affiliation{Dept. of Physics, University of Oxford, Parks Road, Oxford OX1 3PU, UK}
\affiliation{Dept. of Physics, Drexel University, 3141 Chestnut Street, Philadelphia, PA 19104, USA}
\affiliation{Physics Department, South Dakota School of Mines and Technology, Rapid City, SD 57701, USA}
\affiliation{Dept. of Physics, University of Wisconsin, River Falls, WI 54022, USA}
\affiliation{Dept. of Physics and Astronomy, University of Rochester, Rochester, NY 14627, USA}
\affiliation{Department of Physics and Astronomy, University of Utah, Salt Lake City, UT 84112, USA}
\affiliation{Oskar Klein Centre and Dept. of Physics, Stockholm University, SE-10691 Stockholm, Sweden}
\affiliation{Dept. of Physics and Astronomy, Stony Brook University, Stony Brook, NY 11794-3800, USA}
\affiliation{Dept. of Physics, Sungkyunkwan University, Suwon 16419, Korea}
\affiliation{Institute of Basic Science, Sungkyunkwan University, Suwon 16419, Korea}
\affiliation{Dept. of Physics and Astronomy, University of Alabama, Tuscaloosa, AL 35487, USA}
\affiliation{Dept. of Astronomy and Astrophysics, Pennsylvania State University, University Park, PA 16802, USA}
\affiliation{Dept. of Physics, Pennsylvania State University, University Park, PA 16802, USA}
\affiliation{Dept. of Physics and Astronomy, Uppsala University, Box 516, S-75120 Uppsala, Sweden}
\affiliation{Dept. of Physics, University of Wuppertal, D-42119 Wuppertal, Germany}
\affiliation{DESY, D-15738 Zeuthen, Germany}

\author[0000-0001-6141-4205]{R. Abbasi}
\affiliation{Department of Physics, Loyola University Chicago, Chicago, IL 60660, USA}

\author[0000-0001-8952-588X]{M. Ackermann}
\affiliation{DESY, D-15738 Zeuthen, Germany}

\author{J. Adams}
\affiliation{Dept. of Physics and Astronomy, University of Canterbury, Private Bag 4800, Christchurch, New Zealand}

\author[0000-0003-2252-9514]{J. A. Aguilar}
\affiliation{Universit{\'e} Libre de Bruxelles, Science Faculty CP230, B-1050 Brussels, Belgium}

\author[0000-0003-0709-5631]{M. Ahlers}
\affiliation{Niels Bohr Institute, University of Copenhagen, DK-2100 Copenhagen, Denmark}

\author{M. Ahrens}
\affiliation{Oskar Klein Centre and Dept. of Physics, Stockholm University, SE-10691 Stockholm, Sweden}

\author{C. Alispach}
\affiliation{D{\'e}partement de physique nucl{\'e}aire et corpusculaire, Universit{\'e} de Gen{\`e}ve, CH-1211 Gen{\`e}ve, Switzerland}

\author{A. A. Alves Jr.}
\affiliation{Karlsruhe Institute of Technology, Institute for Astroparticle Physics, D-76021 Karlsruhe, Germany }

\author{N. M. Amin}
\affiliation{Bartol Research Institute and Dept. of Physics and Astronomy, University of Delaware, Newark, DE 19716, USA}

\author{R. An}
\affiliation{Department of Physics and Laboratory for Particle Physics and Cosmology, Harvard University, Cambridge, MA 02138, USA}

% ...
\date{\today}
\collaboration{1000}{IceCube Collaboration}

\begin{abstract}
A recent time-integrated analysis of a catalog of 110 candidate neutrino sources revealed a cumulative neutrino excess in the data collected by IceCube between April 6, 2008 and July 10, 2018. This excess, inconsistent with the background hypothesis in the Northern hemisphere at the $3.3~\sigma$ level, is associated with four sources: NGC 1068, TXS 0506+056, PKS 1424+240 and GB6 J1542+6129. This letter presents two time-dependent neutrino emission searches on the same data sample and catalog: a point-source search that looks for the most significant time-dependent source of the catalog by combining space, energy and time information of the events, and a population test based on binomial statistics that looks for a cumulative time-dependent neutrino excess from a subset of sources. Compared to previous time-dependent searches, these analyses enable a feature to possibly find multiple flares from a single direction with an unbinned maximum-likelihood method. M87 is found to be the most significant time-dependent source of this catalog at the level of $1.7~\sigma$ post-trial, and TXS 0506+056 is the only source for which two flares are reconstructed. The binomial test reports a cumulative time-dependent neutrino excess in the Northern hemisphere at the level of $3.0~\sigma$ associated with four sources: M87, TXS 0506+056, GB6 J1542+6129 and NGC 1068.
\end{abstract}



\section{Introduction}
\label{sec:intro}

After more than 100 years since their discovery, the origin and acceleration processes of cosmic rays (CRs) remain unsolved. Relevant hints exist, one being provided by a neutrino event detected by IceCube with most probable energy of 290 TeV which triggered follow-up gamma-ray observations ~\citep{IceCube:2018dnn}. These observations identified in the 50\% containment region for the arrival direction of the IceCube event a classified BL Lac object, though possibly a Flat-Spectrum Radio Quasar (FSRQ) \citep{Padovani:2019xcv}, at redshift $z = 0.34$, known as TXS 0506+056. It was in a flaring state \citep{IceCube:2018dnn} with a chance correlation between the neutrino event and the photon counterpart rejected at the $3~\sigma$ level. The intriguing aspect of the possible coincidence between the neutrino event and the gamma-ray flare hints at TXS 0506+056 being a potential CR source. Additionally, in the analysis of the data prior to the event alert IceCube found a neutrino flare of 110 day duration between 2014/2015~\citep{IceCube:2018cha} at a significance of $3.7~\sigma$, if a Gaussian time window is assumed. In this case, no clear flare has been identified in available gamma-ray data from TXS 0506+056~\citep{Aartsen:2019gxs,Glauch:2019emd}. 
The total contribution of the observed TXS 0506+056 neutrino flares to the diffuse astrophysical flux observed by IceCube~\citep{Aartsen:2013jdh,Aartsen:2014gkd,Aartsen:2016xlq,Aartsen:2017mau} is at most a few percent~\citep{IceCube:2018cha}.
In addition, time-integrated upper limits on stacked catalogs of classes of sources (e.g. tidal disruption events \citep{Stein:2019ivm}, blazars \citep{Aartsen:2016lir}, gamma ray bursts \citep{Aartsen:2017wea}, compact binary mergers \citep{Aartsen:2020mla} and pulsar wind nebulae \citep{Aartsen:2020eof}), 
constrain their contribution to the measured diffuse flux. While these limits depend on assumptions on the emission of such classes of sources, such as their spectral shapes and their uniformity within the class, they indicate that there might be a mixture of contributing classes and still unidentified contributors.

Recently, IceCube performed another analysis on neutrino sources: a time-integrated search for point-like neutrino source signals using ten years of data~\citep{Aartsen:2019fau}. This search uses a maximum-likelihood (ML) method to test the locations of a catalog of 110 selected sources and the full sky. As an intriguing coincidence, the two searches find the hottest spot to be a region including the Seyfert II galaxy NGC 1068, with a significance reported from the catalog search of $2.9~\sigma$. Additionally, a population study of the catalog revealed a $3.3~\sigma$-level incompatibility of the neutrino events from the directions of four Northern sources with respect to the estimated background: NGC 1068, TXS 0506+056, PKS 1424+240 and GB6 J1542+6129.

To fully investigate this catalog of sources, this letter shows the results of a complementary time-dependent study. Time-dependent searches are particularly interesting not only because of their better sensitivity to time-integrated searches for flares of duration $\lesssim 200$ d, due to the suppression of the time-constant background of atmospheric neutrinos, but also because flare events are particularly suitable periods for neutrino production in blazars. In fact, the injection rate of accelerated protons and the density of target photon fields for photo-meson interactions can be noticeably increased during flaring periods of blazars, leading to an enhanced neutrino luminosity $L_\nu \propto L^{1.5\text{--}2}_\gamma$ (see \cite{Zhang:2019htg} and references therein), where $L_\gamma$ is the photon luminosity. 
Apart from the aforementioned evidence of the 2014/2015 flare from the direction of TXS 0506+056, other IceCube time-dependent searches did not find any significant excess. Nevertheless, they constrained specific emission models \citep{Abbasi:2020dfi} or set upper limits on the neutrino emission from selected sources \citep{Aartsen:2015wto}. Triggered searches adopt lightcurves or flare directions from gamma-ray experiments, while sky scans search for largest flares anywhere in the sky.
In this paper, we extend these searches to a multiple flare scan based on a ML method. 
\section{Apparatus and data sample}
\label{sec:detector}
The IceCube Neutrino Observatory~\citep{Achterberg:2006md} is a 1~km$^3$-sized neutrino telescope optimized for detection of high-energy neutrinos above $\sim 100$~GeV. It is located at the South Pole at a depth from $\sim 1.5$ to 2.5~km in the Antarctic ice. It consists
of 86 strings instrumented with 5,160 digital optical modules (DOMs), each equipped with a 10-inch photomultiplier tube~\citep{Abbasi:2010vc} in a pressure-resistant sphere with associated digitizing electronics~\citep{Abbasi:2008aa}.

The dataset used in this analysis comprises 10 years of IceCube data, from 
April 6, 2008 to July 10, 2018~\citep{Aartsen:2019fau,Abbasi:2021bvk}. It includes data from partially-built detector configurations, with 40, 59, 79 strings (IC40, IC59, IC79) described in ~\cite{Abbasi:2010rd,Aartsen:2013uuv,
Aartsen:2016oji}, and from the full detector configuration of 86 strings (IC86-I), described
in~\cite{Aartsen:2014cva}. An updated selection of track-like events that reduces the atmospheric background, described in \cite{Aartsen:2019fau}, is applied to the data recorded in 2012-2018. It leads to an all-sky event rate of $\sim 4$~mHz, dominated by muons from interactions of atmospheric neutrinos from the Northern hemisphere (up-going region, declination $\delta \geqslant -5^{\circ}$) and by high-energy, well-reconstructed atmospheric muons from the Southern hemisphere (down-going region, $\delta < -5^{\circ}$).
The resulting median angular resolution (the difference between the reconstructed simulated event direction and the true primary neutrino) is smaller than $0.60^{\circ}$ above 10~TeV, where it shows an improvement of $\sim 10\%$ with respect to the previous event selection~\citep{Aartsen:2016oji}.

\begin{figure}[hbt]
    \centering
    \includegraphics[width=0.8\textwidth]{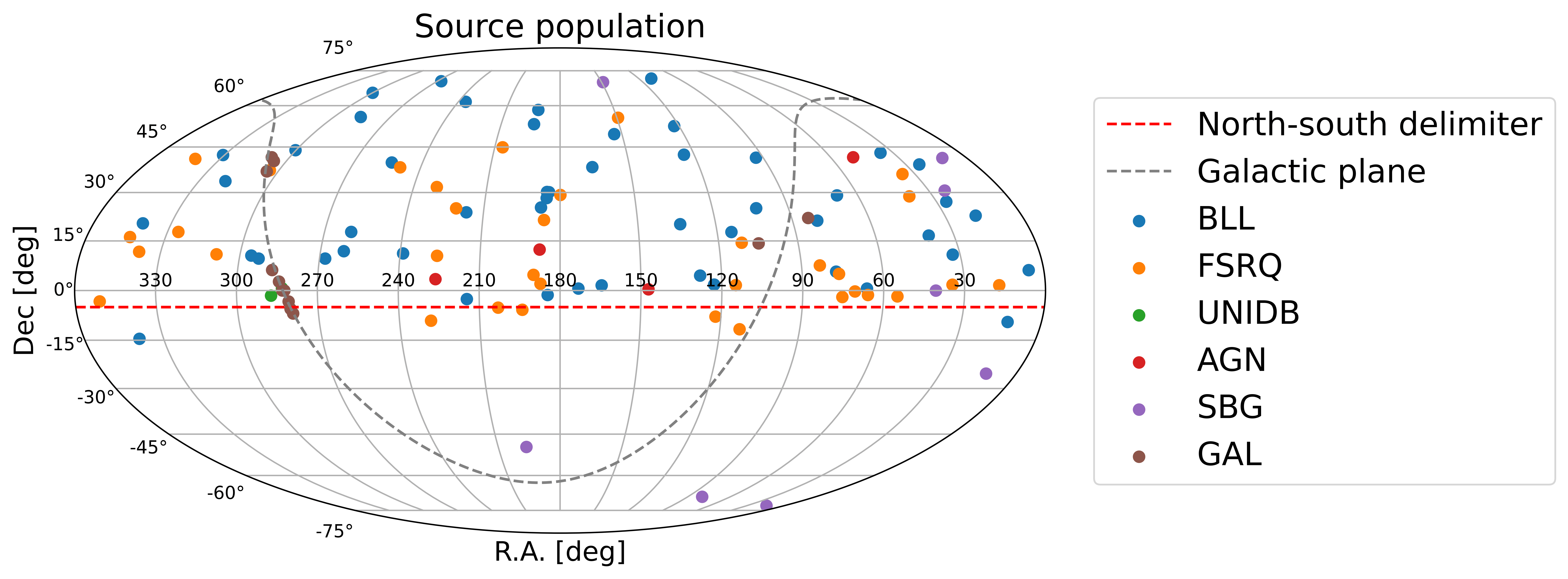}
    \caption{Distribution of the sources in the catalog in equatorial coordinates. They are classified as BL Lacs (BLLs), Flat-Spectrum Radio Quasars (FSRQs), Active Galactic Nuclei (AGNs), Starburst Galaxies (SBGs), Unidentified Blazars (UNIDBs) and galactic sources (GALs). The red line divides the Northern hemisphere (up-going region) and the Southern hemisphere (downgoing region) at declination $\delta=-5^\circ$, where the background is substantially different.}
    \label{fig:skymap}
\end{figure}

The selection of the candidate neutrino sources used in these analyses is described in~\cite{Aartsen:2019fau}. The catalog is composed of extragalactic sources selected from the \textit{Fermi}-LAT 4FGL catalog~\citep{Fermi-LAT:2019yla} that include eight starburst galaxies detected by \textit{Fermi}-LAT, which may host hadronic interactions of cosmic rays with ambient matter, and galactic sources from TeVCat~\citep{Wakely:2007qpa} and gammaCat~\citep{gammaCat:2018}. The catalog is aimed at maximizing the detection probability of IceCube, while considering relevant GeV-TeV information from gamma-ray experiments. In total, 110 sources are selected (see Fig.~\ref{fig:skymap}), 97 located in the Northern hemisphere and 13 in the Southern one. 

\section{Data analysis methods}
\label{sec:analysis}

The presented analyses are based on an unbinned ML method similar to previous IceCube analyses, extended to allow the detection of multiple flares and to handle different IceCube samples (IC40, IC59, IC79, IC86-I, IC86-II-VII) with different detector configurations. Since each IceCube sample is independent, the total 10-year likelihood $\mathcal{L}$ is defined as the product of the likelihoods of each single IceCube sample $\mathcal{L}_j$:
\begin{equation}
    \label{eq:10-year-likelihood}
    \mathcal{L}(\vec{n}_s, \vec{\gamma}, \vec{t}_0, \vec{\sigma}_T)=\prod_{j=\mathrm{sample}}\mathcal{L}_j(\vec{n}_{s,j}, \vec{\gamma}, \vec{t}_0, \vec{\sigma}_T),
\end{equation}
where $\mathcal{L}_j$ is defined as

\begin{equation}
\label{eq:multi-likelihood}
\mathcal{L}_j(\vec{n}_{s,j}, \vec{\gamma}, \vec{t}_0, \vec{\sigma}_T) = \prod_{i=1}^{N_j}\left[\frac{\sum_{f=\mathrm{flares}}n_{s,j}^f\mathcal{S}_j(|\vect{x_s}-\vect{x_i}|,\sigma_i, E_i,t_i; \gamma^f, t_0^f, \sigma_T^f)}{N_j}+\left(1-\frac{\sum_fn_{s,j}^f}{N_j}\right)\mathcal{B}_j(\sin\delta_i, E_i)\right] .
\end{equation}

For each flare $f$, the likelihood in Eq.~\ref{eq:10-year-likelihood} is a function of four parameters described below: the total number of signal-like events in the flare $n_s^f$, the flare spectral index $\gamma^f$, the flaring time $t_0^f$ and the flare duration $\sigma_T^f$. They are denoted with an arrow in the likelihood arguments to indicate that there are as many sets of these four parameters as the number of flares. For each flare $f$, $n_{s,j}^f$ in Eq.~\ref{eq:multi-likelihood} denotes the partial contribution of the $j$-th sample to the total number of signal-like events in that flare, such that $n_s^f=\sum_j n_{s,j}^f$. Such partial contribution $n_{s,j}^f$ is estimated from the relative effective area of the IceCube configuration of the $j$-th sample (determined by Monte Carlo simulations of the detector and varying with spectral index and declination) and the fraction of time that the $f$-th flare stretches on the data-taking period of the $j$-th sample.

For each IceCube sample $j$, with $N_j$ total events, the likelihood in Eq.~\ref{eq:multi-likelihood} is constructed from a single-flare signal probability density function (PDF) $\mathcal{S}_j$, weighted by $n_{s,j}^f$ and summed over all flares from a source (multi-flare signal PDF), and a background PDF $\mathcal{B}_j$. The single-flare signal PDF and the background PDF are the product of a space, energy and time PDFs, as also described in \cite{Aartsen:2015wto}. The spatial signal PDF assumes a cluster of events distributed according to a 2D Gaussian around the source position $\vect{x_s}$, with $\sigma_i$ being the estimated angular uncertainty on the $\vect{x_i}$ position of the $i$-th event. For the signal energy PDF, that depends on the declination $\delta_i$ and the energy proxy $E_i$ of the events (the energy as measured by IceCube from visible light released in the detector by muon tracks), an unbroken power law $\propto E^{-\gamma^f}$ is used. The spectral index $\gamma^f$ is bound within $1\le\gamma^f\le4$ and can be different for each flare $f$. The signal time PDF of each flare $f$ is provided by a one-dimensional Gaussian $\propto \exp{[-(t_i-t_0^f)^2/(2\sigma_T^{f2})]}$, where $t_i$ is the time of the $i$-th event. Its normalization is such that the integral of the time PDF across the up times of each IceCube sample is 1. The central time of each Gaussian flare $t_0^f$ is constrained within the 10-year period of the analyzed data and the flare duration $\sigma_T^f$ cannot exceed an upper limit of 200 days, above which time-integrated searches are more sensitive than time-dependent ones. For computational efficiency, the signal time PDF of each flare is truncated at $\pm 4\sigma_T^f$, where the flare can be considered concluded.

The spatial background PDF is obtained through a data-driven method by scrambling the time of the events and correcting the right ascension accordingly, assuming fixed local coordinates (azimuth, zenith). It depends only on the declination $\delta_i$ of the events and it is uniform in right ascension. Due to the natural tendency of the reconstruction to be more efficient if the direction of the source is aligned with the strings of the detector, an azimuth-dependent correction is applied to the spatial background PDF. Such correction is relevant for time scales shorter than one day, whereas it is negligible for longer time scales, since any azimuth dependency is averaged out by the Earth rotation. The background energy PDF is taken from scrambled data as well, and it is fully described in~\cite{Aartsen:2013uuv}. It depends on the declination $\delta_i$ and the energy proxy $E_i$ of the events. The background time PDF is uniform, as expected for atmospheric muons and neutrinos if seasonal sinusoidal variations are neglected. The maximal amplitude for these variations is 10\% for the downgoing muons produced in the polar atmosphere  and smaller for atmospheric neutrinos coming from all latitudes \citep{Gaisser:2013lrk}.

The test statistic (TS) is defined as:
\begin{equation}
\label{eq:teststatistic}
\mathrm{TS}=-2\ln\left[\frac{1}{2}\left(\prod_{f=\mathrm{flares}}\frac{T_{live}}{\hat{\sigma}_T^f I\left[\hat{t}_0^f, \hat{\sigma}_T^f\right]}\right)\times\frac{\mathcal{L}(\vec{n}_s=\vec{0})}{\mathcal{L}(\vec{\hat{n}}_s, \vec{\hat{\gamma}}, \vec{\hat{t}}_0, \vec{\hat{\sigma}}_T)}\right] ,
\end{equation} 
where the parameters that maximize the likelihood function in Eq.~\ref{eq:10-year-likelihood} are denoted with a hat and $\mathcal{L}(\vec{n}_s=\vec{0})$ is the background likelihood, obtained from Eq.~\ref{eq:10-year-likelihood} by setting $n_s^f=0$ for all the flares.
The likelihood ratio is multiplied by a marginalization term intended to penalize short flares, similarly used in previous time-dependent single-flare IceCube analyses to correct a natural bias of the likelihood towards selecting short flares. This was discussed in~\cite{Braun:2009wp} for the single-flare analysis. For the multi-flare analysis, the numerical factor $1/2$ in the equation above is chosen such that the marginalization term has the same form as the single-flare one when the true hypothesis is a single flare. The factor $0<I\left[\hat{t}_0^f, \hat{\sigma}_T^f\right]<1$ is defined as
\begin{equation}
I\left[\hat{t}_0^f, \hat{\sigma}_T^f\right]=\int_{T_{live}}\frac{1}{\sqrt{2\pi}\sigma_T^f}\exp{\left[-\frac{(t-t_0^f)^2}{2\sigma_T^{f2}}\right]}dt ,
\end{equation}
where $T_{live}$ is the full period of the analysis. It is introduced in this analysis to correct for boundary effects of Gaussian-modeled time-dependent searches.

A pre-selection of interesting time frames is made by evaluating the TS of the events, looking for clusters of high-energy events in time and space, and imposing that $\mathrm{TS} \ge 2$.
This value is chosen to reduce the reconstruction probability of `fake' multiple flares to $\lesssim0.1\%$ under the null hypothesis (see Appendix~\ref{sec:multi-flare_algorithm}). Clusters of events fulfilling the above requirement are called candidate flares. For the candidate flares with central times $t_0^f$ overlapping within $\pm 4\sigma^f_T$, only the one with the highest TS is retained. If no candidate flares are found, the pre-selection reports as candidate flare the one cluster with the highest TS. The parameters of these candidate flares are used as seed values for the ML fit.

The first analysis presented in this letter, the point source search, looks for the most significant flaring source in each hemisphere. To do so, the TS is maximized at the location of each source and converted into a local pre-trial p-value $p_{loc}$ by comparing the observed result with a distribution of TS produced under the null hypothesis ($\vec{n}_s=\vec{0}$). The source corresponding to the lowest pre-trial p-value in the respective hemisphere is reported, together with its post-trial p-value after accounting for the number of trials, i.e. often called look-elsewhere effect. The post-trial p-value is estimated in each hemisphere by repeating the analysis on background scrambles, picking up the lowest p-value in each scramble and counting the fraction of such background p-values that are smaller than the lowest pre-trial p-value observed in the data.

The second analysis presented in this paper, a population study, is a binomial test that looks for an excess of neutrino emission from a subset of sources that are not strong enough to emerge individually. It aims at determining whether the pre-trial p-values observed in a particular hemisphere are compatible as a whole with a background scenario. The p-values of the sources are ranked from the lowest to the highest, and in each hemisphere the pre-trial binomial p-value $P_{bin}(k)$ is calculated as a function of the source index $k$ \cite{OSullivan:2019rpq,Aartsen:2019fau}:
\begin{equation}
P_{bin}(k) = \sum_{m=k}^N \frac{N!}{(N-m)!\,m!}p_k^m(1-p_k)^{N-m} .
\end{equation}

Here $p_k$ is the pre-trial p-value of the $k$-th source in the sorted list of $N$ sources ($N=97$ in the North, $N=13$ in the South). After scanning all the source indices $k$, the smallest binomial p-value is corrected for trials and reported as post-trial in each hemisphere, together with the corresponding number of sources. The post-trial binomial p-value is estimated in each hemisphere by producing many background realizations of the catalog, picking up the smallest binomial p-value in each background realization and counting the fraction of such background binomial p-values that are smaller than the binomial p-value observed in the data.
\section{Results} \label{sec:results}

\begin{figure}[htbp]
  \centering
  \includegraphics[width=0.9\linewidth]{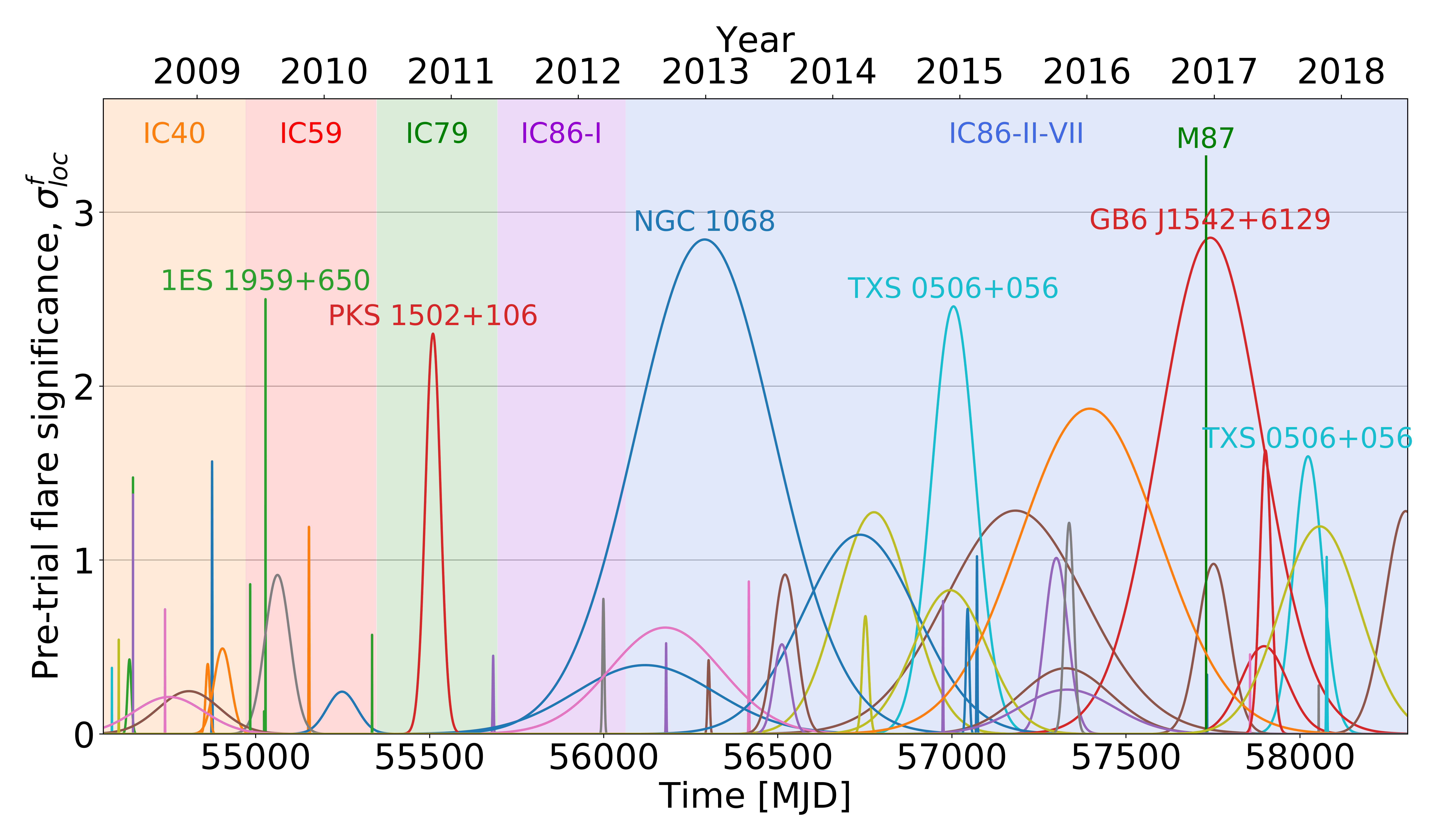} 
  \caption{Pre-trial flare significance $\sigma_{loc}^f$ for the sources of the catalog. For all sources a single flare has been found, except for TXS 0506+056 for which 2 flares are found. In this case, the pre-trial significance of each individual flare is calculated as described in Appendix \ref{sec:singleflare_significance}. The sources of the catalog with multi-flare pre-trial significance $\sigma_{loc}\ge2$ are labeled with their names.}
  
  \label{fig:best_fit_flares}
\end{figure}

The point-source search identifies M87 as the most significant source in the Northern hemisphere, with a pre-trial p-value of $p_{loc}=4.6\times10^{-4}$, which becomes $4.3\times10^{-2}$ ($1.7~\sigma$) post-trial. In the Southern hemisphere, the most significant source is PKS 2233-148 with a pre-trial p-value of $p_{loc}=0.092$ and post-trial p-value of $0.72$. TXS 0506+056 is the only source of the catalog for which 2 flares are found. The time profiles of the neutrino flares reconstructed by this analysis at the location of each source, together with their pre-trial significance $\sigma_{loc}^f$, are visualized in Fig.~\ref{fig:best_fit_flares}. For the sake of clarity, the flare significance is denoted as $\sigma_{loc}^f$ while the overall multi-flare significance is referred to as $\sigma_{loc}=\sqrt{\sum_f\sigma_{loc}^{f2}}$. For single-flare sources (all but TXS 0506+056) the flare and multi-flare significances coincide. 

\begin{figure}[htbp]
  \centering
  \includegraphics[width=.49\linewidth]{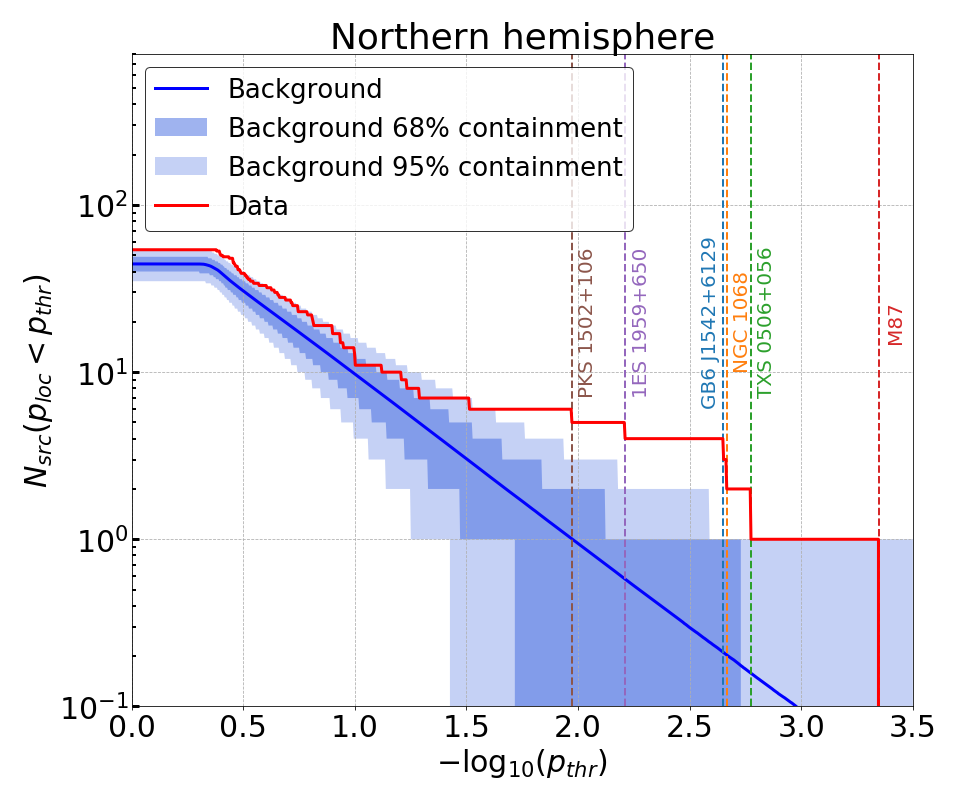}  
  \includegraphics[width=.49\linewidth]{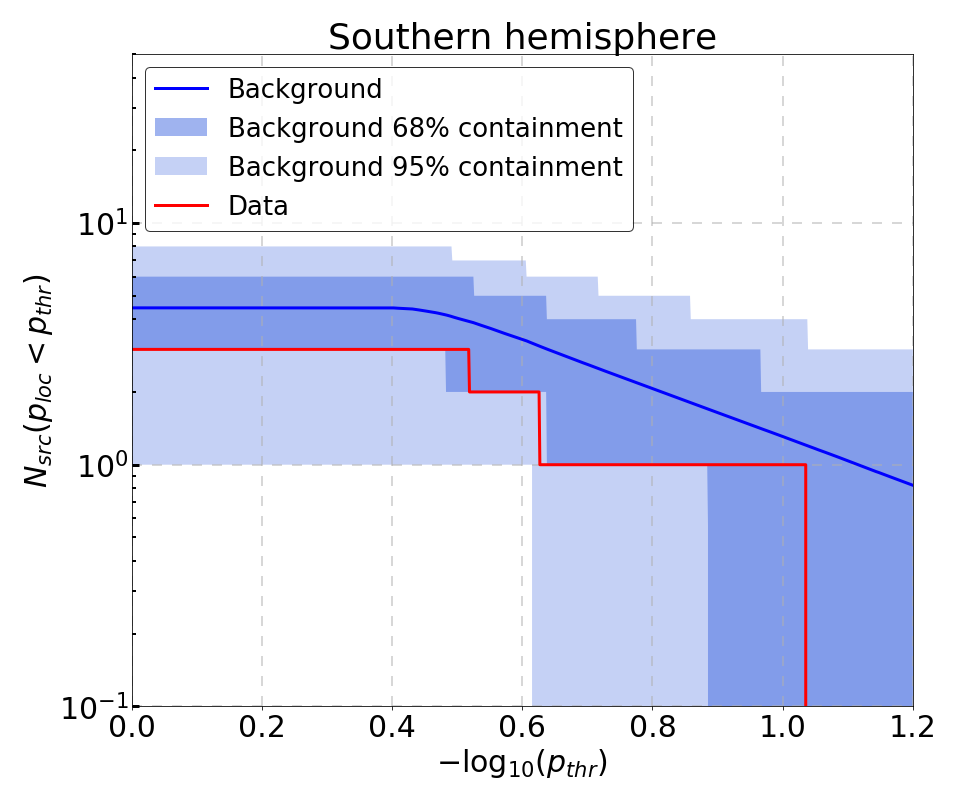}  
  \caption{Cumulative distributions of the pre-trial p-values of the sources of the catalog in the Northern (left) and Southern (right) hemispheres. The cumulative p-values of the unblinded data are shown in red and compared to the background expectations in blue.}
  \label{fig:pvalues_distribution}
\end{figure}

The cumulative distributions of pre-trial p-values at the location of the sources of the catalog, used as inputs to the population study, are shown in Fig.~\ref{fig:pvalues_distribution}.

\begin{figure}[htbp]
  \centering
  \includegraphics[width=.47\linewidth]{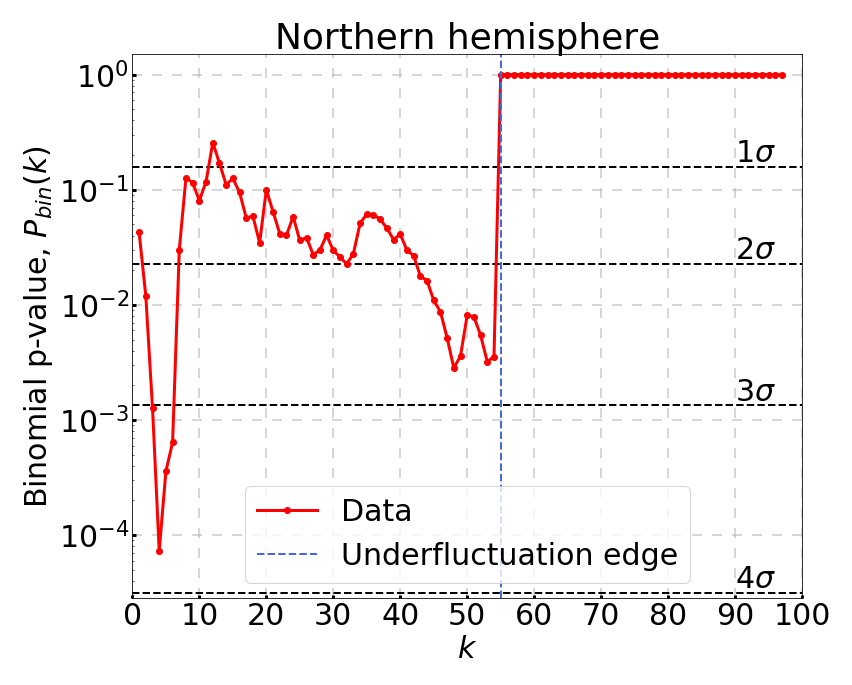}  
  \includegraphics[width=.49\linewidth]{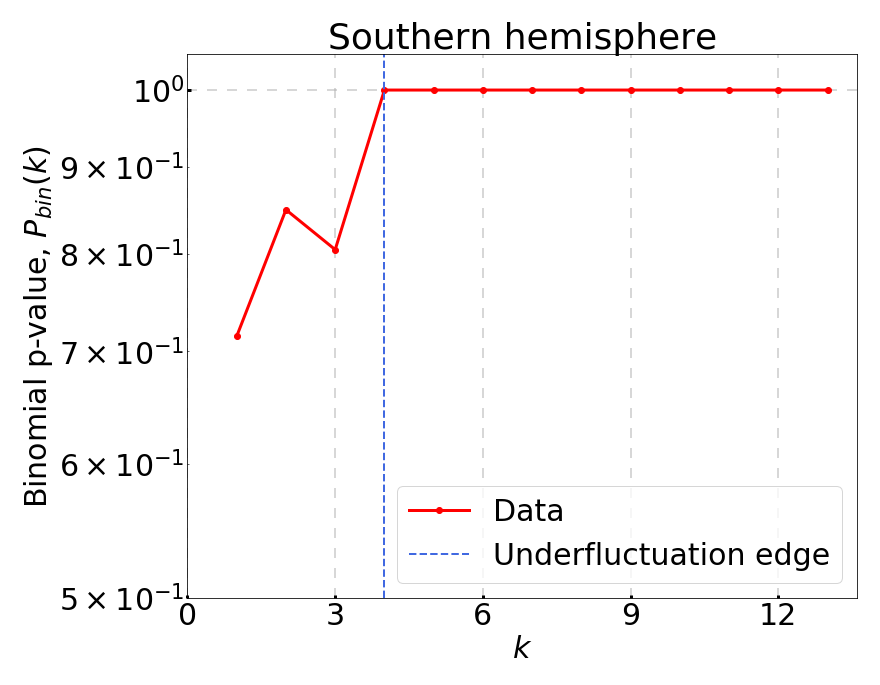}   
  \caption{Pre-trial binomial p-value $P_{bin}(k)$ as a function of the source index $k$ in the Northern (left) and Southern (right) hemispheres. The edge with the under-fluctuating sources, with binomial p-value set to 1, is shown in blue.}
  \label{fig:binomial_test}
\end{figure}

The pre-trial binomial p-value is shown in Fig.~\ref{fig:binomial_test} as a function of the source index $k$. The smallest binomial p-value is selected in each hemisphere and converted into a post-trial binomial p-value as described in Section~\ref{sec:analysis}. In the Northern hemisphere the smallest pre-trial binomial p-value is $7.3\times10^{-5}$ ($3.8~\sigma$) when $k=4$ sources are considered (M87, TXS 0506+056, GB6 J1542+6129, NGC 1068), corresponding to a post-trial p-value of $1.6\times 10^{-3}$ ($3.0~\sigma$). In the Southern hemisphere the smallest pre-trial binomial p-value is 0.71, obtained by $k=1$ source (PKS 2233-148) and corresponding to a post-trial p-value of $0.89$.

The results of the two searches are summarized in Table~\ref{tab:summary_results}. Having not found any significant time-dependent excess, upper limits on the neutrino emission from the sources of the catalog are estimated as discussed in Appendix~\ref{sec:sens_DP_upLims}, using Eq.~\ref{eq:time-integrated_flux} and~\ref{eq:flux_definition}.

\begin{table}
\centering
\begin{tabular}{>{\centering\arraybackslash}m{2.8cm} >{\centering\arraybackslash}m{2.5cm} >{\centering\arraybackslash}m{1.7cm} >{\centering\arraybackslash}m{2.5cm}}
    \multicolumn{4}{c}{Summary of the results}\\
    \hline
    \hline
    \multirow{2}{*}{Analysis} & \multirow{2}{*}{Hemisphere} & \multicolumn{2}{c}{p-value}\\ & & Pre-trial & Post-trial \\[3pt] \hline
    \multirow{2}{*}{Point-source} & North & $4.6\times10^{-4}$ & $4.3\times10^{-2}$ ($1.7~\sigma$)\\ & South & $9.2\times 10^{-2}$ & 0.72\\  [3pt] \hline
    \multirow{2}{*}{Binomial test} & North & $7.3\times10^{-5}$ & $1.6\times10^{-3}$ ($3.0~\sigma$) \\ & South & $0.71$ & $0.89$\\
    \hline
    \hline
\end{tabular}
\caption{Summary of the results of the two analyses: for the point-source search the results of the best sources in the Northern (M87) and Southern (PKS 2233-148) hemisphere are reported.}
\label{tab:summary_results}
\end{table}
\section{Conclusions} \label{sec:conclusions}

The time-dependent point-source search presented in this letter identified M87 as the most significant source in the Northern hemisphere, with $\hat{n}_s=3$ signal-like neutrino events in a time window of $\hat{\sigma}_T=2.0$ minutes and with a soft spectrum ($\hat{\gamma}=3.95$). The post-trial significance of M87 is found to be $1.7~\sigma$. Because of the quite short time lag between the events, the time-dependent search is more sensitive than the time-integrated one, which explains the absence of significant signals in previous IceCube time-integrated analyses that had included M87. For the case of~\cite{OSullivan:2019rpq}, a smaller data sample from Apr. 26, 2012 to May 11, 2017 was used. The difference in significance is due to small changes in the event reconstruction and angular uncertainty estimation between the two samples.

This analysis also identifies the two known flares at the location of TXS 0506+056, one corresponding to the most significant flare at $\sim 57000$ MJD \citep{IceCube:2018cha} and the other related to the high-energy event alert IceCube-170922A detected on 22 Sep. 2017 \citep{IceCube:2018dnn}. Although these two flares are consistently identified, the significance of the result at the location of TXS 0506+056 is lower than the one reported in~\citep{IceCube:2018cha}. This is due to the new data selection \citep{Abbasi:2021bvk} described in Section~\ref{sec:detector}, which introduces a different energy reconstruction from the past one~\citep{Abbasi:2021bvk}. Further information about the reduced significance of TXS 0506+056 resulting from this analysis are provided in Appendix~\ref{sec:TXS_significance_investigation}.



The time-dependent binomial test of the Northern hemisphere suggests an incompatibility at $3.0~\sigma$ significance of the neutrino events from four sources with respect to the overall Northern background expectation. Of the four most significant sources in the Northern hemisphere, three are common with the time-integrated analysis~\citep{Aartsen:2019fau}, namely NGC 1068, TXS 0506+056, GB6 J1542+6129, whereas a fourth source (M87) is different  and shows a strong time-dependent behavior. However, the results of the time-dependent and time-integrated binomial test partly overlap, as both share the same space and energy PDFs in the likelihood definition in Eq.~\ref{eq:multi-likelihood} and both select the same three out of four sources. For this reason, although a time-dependent structure of the data is suggested by the binomial test, a time-independent scenario cannot be excluded by this analysis (see Appendix~\ref{app:variab} for a further discussion).


No significant result is found in the Southern hemisphere. This is consistent with the lower sensitivity due to the substantially larger background of atmospheric muons in the Southern hemisphere.
\section*{Acknowledgements}
The IceCube collaboration gratefully acknowledges the support from the following agencies and institutions: USA {\textendash} U.S. National Science Foundation-Office of Polar Programs,
U.S. National Science Foundation-Physics Division,
U.S. National Science Foundation-EPSCoR,
Wisconsin Alumni Research Foundation,
Center for High Throughput Computing (CHTC) at the University of Wisconsin{\textendash}Madison,
Open Science Grid (OSG),
Extreme Science and Engineering Discovery Environment (XSEDE),
Frontera computing project at the Texas Advanced Computing Center,
U.S. Department of Energy-National Energy Research Scientific Computing Center,
Particle astrophysics research computing center at the University of Maryland,
Institute for Cyber-Enabled Research at Michigan State University,
and Astroparticle physics computational facility at Marquette University;
Belgium {\textendash} Funds for Scientific Research (FRS-FNRS and FWO),
FWO Odysseus and Big Science programmes,
and Belgian Federal Science Policy Office (Belspo);
Germany {\textendash} Bundesministerium f{\"u}r Bildung und Forschung (BMBF),
Deutsche Forschungsgemeinschaft (DFG),
Helmholtz Alliance for Astroparticle Physics (HAP),
Initiative and Networking Fund of the Helmholtz Association,
Deutsches Elektronen Synchrotron (DESY),
and High Performance Computing cluster of the RWTH Aachen;
Sweden {\textendash} Swedish Research Council,
Swedish Polar Research Secretariat,
Swedish National Infrastructure for Computing (SNIC),
and Knut and Alice Wallenberg Foundation;
Australia {\textendash} Australian Research Council;
Canada {\textendash} Natural Sciences and Engineering Research Council of Canada,
Calcul Qu{\'e}bec, Compute Ontario, Canada Foundation for Innovation, WestGrid, and Compute Canada;
Denmark {\textendash} Villum Fonden and Carlsberg Foundation;
New Zealand {\textendash} Marsden Fund;
Japan {\textendash} Japan Society for Promotion of Science (JSPS)
and Institute for Global Prominent Research (IGPR) of Chiba University;
Korea {\textendash} National Research Foundation of Korea (NRF);
Switzerland {\textendash} Swiss National Science Foundation (SNSF);
United Kingdom {\textendash} Department of Physics, University of Oxford.

\appendix
\section{Sensitivity, discovery potential and upper limits}
\label{sec:sens_DP_upLims}

\begin{figure}[ht]
	\centering
	\includegraphics[width=.49\linewidth]{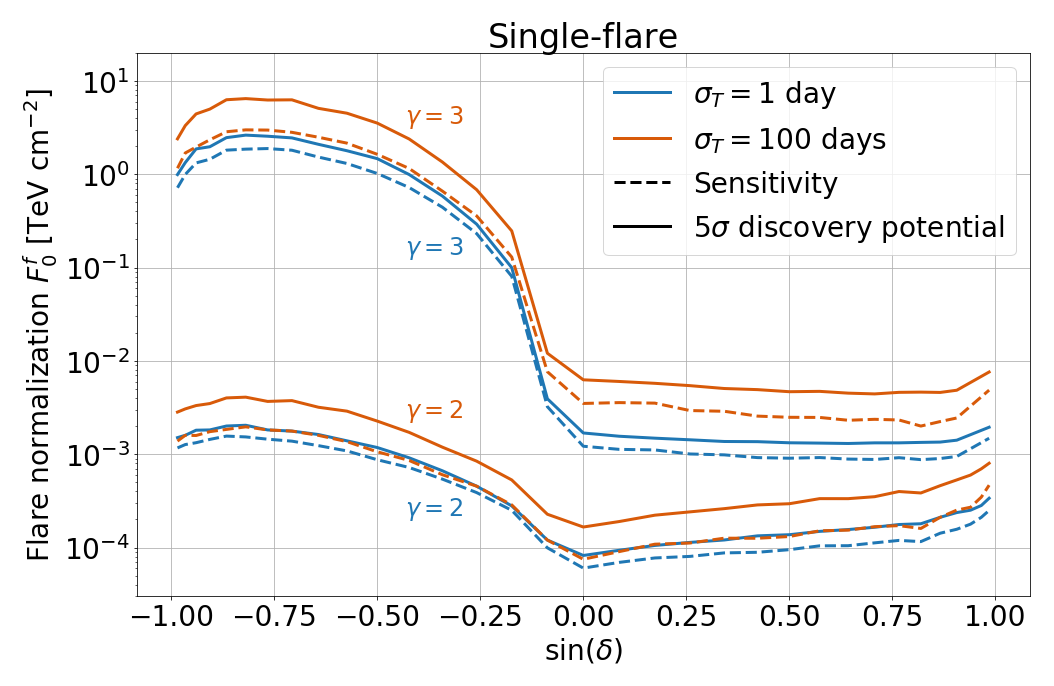} 
	\includegraphics[width=.49\linewidth]{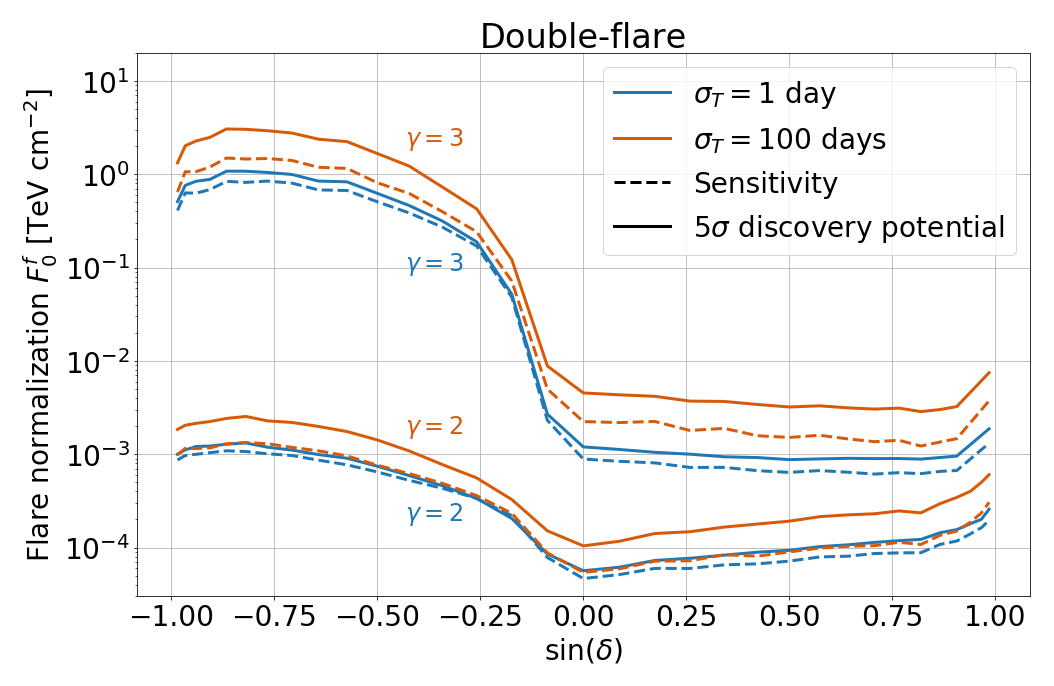} 
	\caption{Sensitivity (dashed lines) and $5~\sigma$ DP (solid lines) of the multi-flare analysis vs declination, expressed in terms of the flux normalization factor per flare $F_0^f$ defined in Eq.~\ref{eq:time-integrated_flux}, under the hypothesis of a single (left plot) and a double (right plot) signal flare. 
	The assumed energy dependence of the flares has a spectral index of $\gamma^f = 2$ and $\gamma^f = 3$ (see labels), and a flare duration of $\sigma_T^f = 1$~day (blue lines) and $\sigma_T^f = 100$~days (orange lines). The double-flare 
	assumes two identical and well separated flares.}
	\label{fig:sensDP}
\end{figure}

The sensitivity and discovery potential (DP) are evaluated by injecting a fake signal in the dataset and looking at the signal-like TS distributions. The sensitivity is defined as the signal flux required such that the resulting TS is greater than the background median in 90\% of the trials; the $5~\sigma$ DP is defined as the signal flux required such that the resulting TS is greater than the $5~\sigma$ threshold of the background TS distribution in 50\% of the trials. The sensitivity and $5~\sigma$ discovery potential (DP) of the multi-flare analysis as a function of the declination are shown in Fig.~\ref{fig:sensDP} for a single (left) and a double (right) signal flare hypothesis. In the latter case, the intensity and spectral shape of the two flares are the same.

The sensitivity and $5~\sigma$ DP are expressed in terms of a time-integrated flux:
\begin{equation}
    \label{eq:time-integrated_flux}
    F = \int_{T_{live}}E^2\Phi(E,t)dt=\sum_{f=\mathrm{flares}}F_0^f\left(\frac{E}{\mathrm{TeV}}\right)^{2-\gamma^f},
\end{equation}
where $F_0^f$ is the time-integrated flux normalization factor of the $f$-th flare, independent of the flare duration $\sigma_T^f$ and carrying the units of an energy divided by an area, and $\Phi(E,t)$ is the overall flux, defined as the sum of the flux of all the flares from a single direction:
\begin{equation}
    \label{eq:flux_definition}
    \Phi(E,t)=\sum_{f=\mathrm{flares}}\frac{F_0^f}{\sqrt{2\pi}\sigma_T^f}\left(\frac{E}{\mathrm{TeV}}\right)^{-\gamma^f}G^f(t|t_0^f,\sigma_T^f).
\end{equation}
In Eq. \ref{eq:flux_definition}, $G^f(t|t_0^f,\sigma_T^f)=\exp\left[-\frac{1}{2}\left(\frac{t-t_0^f}{\sigma_T^f}\right)^2\right]$ is the Gaussian time profile of the $f$-th flare.

Sensitivities and DPs are shown in Fig.~\ref{fig:sensDP} for two different hypotheses of the spectral index of the flares ($\gamma^f=2$ and $\gamma^f=3$) and two different flare durations ($\sigma_T^f=1$ day and $\sigma_T^f=100$ days). In the double-flare case, two identical and well separated flares are assumed, with the same spectral index $\gamma^f$, flare duration $\sigma_T^f$ and time-integrated flux normalization per flare $F_0^f$.

The 90\% confidence level (CL) upper limits on the flux of each source of the catalog are defined as the flux required to produce a TS distribution that exceeds the unblinded TS of the respective source in 90\% of the trials. These upper limits are expressed in terms of a time-integrated flux by mean of the factor $F_{90\%}$, defined as:

\begin{equation}
    \label{eq:time-int_flux_upLims}
    F = F_{90\%}\sum_{f=\mathrm{flares}}\left(\frac{E}{\mathrm{TeV}}\right)^{2-\gamma^f}.
\end{equation}

In the case of TXS 0506+056, the only observed multi-flare source of the catalog, the upper limits are evaluated assuming the same flare intensity for the two flares. As a matter of fact, only one global factor $F_{90\%}$ appears in Eq.~\ref{eq:time-int_flux_upLims}.

The upper limits of the not under-fluctuating sources of the catalog, together with the coordinates, maximum-likelihood parameters and pre-trial p-values, are reported in Table~\ref{tab:PS_results1}. To calculate these upper limits, a spectral index $\gamma^f=2$ in Eq. \ref{eq:time-int_flux_upLims} is assumed for all the flares, whereas the flare time $t_0^f$ and duration $\sigma_T^f$ are taken as the maximum-likelihood parameters. Only one flare is injected for each source, except for TXS 0506+056 for which two flares are injected, according to the maximum-likelihood results.

\setlength\LTleft{-3cm}
\begin{center}
	\begin{longtable}{>{\centering\arraybackslash}m{3.1cm} >{\centering\arraybackslash}m{1.0cm} >{\centering\arraybackslash}m{1.0cm} >{\centering\arraybackslash}m{1.1cm} >{\centering\arraybackslash}m{1.0cm} >{\centering\arraybackslash}m{2.3cm} >{\centering\arraybackslash}m{2.2cm} >{\centering\arraybackslash}m{1.5cm} >{\centering\arraybackslash}m{1.9cm}} 
	\hline
		\multicolumn{9}{|c|}{catalog results}\\ \hline
		Source & R.A. & $\delta$ & $\hat{n}_s$ & $\hat{\gamma}$ & $\hat{t}_0$ & $\hat{\sigma}_T$ & $-\log_{10}(p_{loc})$ & $F_{90\%}\times10^{4}$\\
		 & [ deg ] & [ deg ] & & & [ MJD ] & [ days ] & & [ TeV cm$^{-2}$ ] \\ [3pt]
		 \midrule
		\endfirsthead
		\midrule
		Source & R.A. & $\delta$ & $\hat{n}_s$ & $\hat{\gamma}$ & $\hat{t}_0$ & $\hat{\sigma}_T$ & $-\log_{10}(p_{loc})$  & $F_{90\%}\times10^{4}$\\
		 & [ deg ] & [ deg ] & & & [ MJD ] & [ days ] & & [ TeV cm$^{-2}$ ] \\[3pt]
		\midrule
		\endhead
		S5 0716+71 & 110.49 & 71.34 & -- & -- & -- & -- & -- & --\\
		S4 1749+70 & 267.15 & 70.10 & -- & -- & -- & -- & -- & --\\
		M82 & 148.95 & 69.67 & 27.8 &4.0 & 57395.8 & 200.0 & 1.51 & 5.7\\
		1ES 1959+650 & 300.01 & 65.15 & 3.9 &3.3 & 55028.4 & $1.8\times10^{-1}$ & 2.21 & 3.8\\
		\textbf{GB6 J1542+6129} & \textbf{235.75} & \textbf{61.50} & $\mathbf{23.7^{+9.7}_{-7.9}}$ & $\mathbf{2.7^{+0.5}_{-0.3}}$ & $\mathbf{57740^{+80}_{-60}}$ & $\mathbf{147^{+110}_{-25}}$ & \textbf{2.67} & \textbf{5.3}\\
		PG 1246+586 & 192.08 & 58.34 & -- & -- & -- & -- & -- & --\\
		TXS 1902+556 & 285.80 & 55.68 & 3.2 &4.0 & 54862.5 & 6.0 & 0.46 & 3.6\\ 
		4C +55.17 & 149.42 & 55.38 & 11.2 &3.6 & 58303.7 & 59.7 & 1.00 & 2.5\\ 
		S4 1250+53 & 193.31 & 53.02 & 6.1 &2.2 & 55062.9 & 35.9 & 0.74 & 3.7\\ 
		1ES 0806+524 & 122.46 & 52.31 & 6.5 &3.1 & 55248.3 & 43.3 & 0.39 & 2.8\\ 
		1H 1013+498 & 153.77 & 49.43 & 3.1 &2.2 & 58053.6 & $2.7\times10^{-1}$ & 0.41 & 1.2\\ 
		B3 1343+451 & 206.40 & 44.88 & 4.0 &2.7 & 57856.5 & $2.8\times10^{-1}$ & 0.49 & 1.2\\ 
		MG4 J200112+4352 & 300.30 & 43.89 & 11.6 &2.0 & 56776.2 & 105.9 & 1.00 & 2.6\\
		3C 66A & 35.67 & 43.04 & -- & -- & -- & -- & -- & --\\
		S4 0814+42 & 124.56 & 42.38 & 3.4 &2.6 & 56301.3 & 3.1 & 0.47 & 1.3\\ 
		BL Lac & 330.69 & 42.28 & 3.8 &4.0 & 54637.6 & 5.6 & 0.48 & 2.5\\
		2HWC J2031+415 & 307.93 & 41.51 & 18.8 & 3.4 & 58056.8 & 114.0 & 0.93 & 2.4\\
		NGC 1275 & 49.96 & 41.51 & -- & -- & -- & -- & -- & --\\
		B3 0609+413 & 93.22 & 41.37 & 8.7 &1.7 & 56736.2 & 163.7 & 0.90 & 2.5\\ 
		M31 & 10.82 & 41.24 & 8.6 &2.3 & 57900.7 & 16.4 & 1.29 & 2.1\\
		TXS 2241+406 & 341.06 & 40.96 & 3.8 &2.9 & 55334.5 & $2.5\times10^{-1}$ & 0.55 & 1.7\\
		Gamma Cygni & 305.56 & 40.26 & 5.8 &1.5 & 57336.8 & 13.0 & 0.95 & 1.8\\
		Mkn 501 & 253.47 & 39.76 & -- & -- & -- & -- & -- & --\\
		B3 0133+388 & 24.14 & 39.10 & -- & -- & -- & -- & -- & --\\
		Mkn 421 & 166.12 & 38.21 & 2.9 &2.1 & 54875.0 & $7.6\times10^{-1}$ & 1.23 & 2.8\\
		4C +38.41 & 248.82 & 38.14 & 6.2 &2.1 & 56751.6 & 9.0 & 0.60 & 1.5\\ 
		MG2 J201534+3710 & 303.92 & 37.19 & 3.9 &1.3 & 57326.7 & 129.4 & 0.45 & 1.8\\ 
		MGRO J2019+37 & 304.85 & 36.80 & 4.2 &1.4 & 57330.9 & 135.0 & 0.40 & 1.7\\
		B2 0218+357 & 35.28 & 35.94 & -- & -- & -- & -- & -- & --\\
		B2 2114+33 & 319.06 & 33.66 & -- & -- & -- & -- & -- & --\\
		B2 1520+31 & 230.55 & 31.74 & 5.0 &2.4 & 55999.0 & 2.7 & 0.66 & 1.2\\
		NGC 598 & 23.52 & 30.62 & 4.9 &1.8 & 56520.7 & 33.0 & 0.75 & 1.7\\
		PG 1218+304 & 185.34 & 30.17 & 2.0 &2.4 & 54647.8 & $2.1\times10^{-2}$ & 1.12 & 2.1\\ 
		B2 1215+30 & 184.48 & 30.12 & 2.0 &2.4 & 54647.8 & $2.1\times10^{-2}$ & 1.21 & 2.2\\
		Ton 599 & 179.88 & 29.24 & 2.0 &1.7 & 55024.2 & $3.0\times10^{-3}$ & 0.45 & 1.2\\
		MG2 J043337+2905 & 68.41 & 29.10 & -- & -- & -- & -- & -- & --\\
		4C +28.07 & 39.48 & 28.80 & -- & -- & -- & -- & -- & --\\
		W Comae & 185.38 & 28.24 & 3.1 &3.4 & 55682.4 & 1.5 & 0.49 & 1.2\\
		TXS 0141+268 & 26.15 & 27.09 & -- & -- & -- & -- & -- & --\\
		ON 246 & 187.56 & 25.30 & -- & -- & -- & -- & -- & --\\
		1ES 0647+250 & 102.70 & 25.06 & -- & -- & -- & -- & -- & --\\
		PKS 1441+25 & 220.99 & 25.03 & 4.1 &1.7 & 56994.7 & 105.6 & 0.69 & 1.8\\ 
		PKS 1424+240 & 216.76 & 23.80 & 17.9 &2.8 & 57182.6 & 200.0 & 1.00 & 2.2\\
		S2 0109+22 & 18.03 & 22.75 & 4.6 &4.0 & 55153.2 & $9.2\times10^{-1}$ & 0.93 & 1.6\\
		Crab nebula & 83.63 & 22.01 & -- & -- & -- & -- & -- & --\\
		4C +21.35 & 186.23 & 21.38 & 2.0 &2.3 & 55690.3 & $1.2\times10^{-3}$ & 0.64 & 0.9\\
		TXS 0518+211 & 80.44 & 21.21 & -- & -- & -- & -- & -- & --\\
		RGB J2243+203 & 340.99 & 20.36 & 11.2 &3.6 & 57300.1 & 33.0 & 0.81 & 1.5\\ 
		OJ 287 & 133.71 & 20.12 & 3.6 &2.6 & 56416.8 & $8.4\times10^{-1}$ & 0.72 & 1.0\\ 
		PKS 1717+177 & 259.81 & 17.75 & 2.0 &3.3 & 54587.2 & $2.0\times10^{-1}$ & 0.45 & 1.3\\ 
		OX 169 & 325.89 & 17.73 & -- & -- & -- & -- & -- & --\\
		PKS 0735+17 & 114.54 & 17.71 & -- & -- & -- & -- & -- & --\\
		PKS 0235+164 & 39.67 & 16.62 & -- & -- & -- & -- & -- & --\\
		3C 454.3 & 343.50 & 16.15 & 5.1 &2.0 & 56119.1 & 200.0 & 0.46 & 1.3\\
		4C +14.23 & 111.33 & 14.42 & 3.1 &2.0 & 58076.6 & 1.2 & 0.81 & 1.0\\
		PSR B0656+14 & 104.95 & 14.24 & -- & -- & -- & -- & -- & --\\
		\textbf{M87} & \textbf{187.71} & \textbf{12.39} & $\mathbf{3.0^{+2.0}_{-1.4}}$ &$\mathbf{4.0^{+0.9}_{-0.9}}$ & $\mathbf{57730.031^{+0.001}_{-0.001}}$ & $\mathbf{1.4^{+1.3}_{-0.4}\times10^{-3}}$ & \textbf{3.35} & \textbf{0.9}\\
		1H 1720+117 & 261.27 & 11.88 & -- & -- & -- & -- & -- & --\\
		CTA 102 & 338.15 & 11.73 & -- & -- & -- & -- & -- & --\\
		PG 1553+113 & 238.93 & 11.19 & -- & -- & -- & -- & -- & --\\
		PKS 2032+107 & 308.85 & 10.94 & -- & -- & -- & -- & -- & --\\
		MG1 J021114+1051 & 32.81 & 10.86 & 2.8 &2.1 & 56179.2 & $8.9\times10^{-1}$ & 0.52 & 0.9\\ 
		1RXS J194246.3+1 & 295.70 & 10.56 & 4.2 &3.4 & 54904.8 & 24.3 & 0.51 & 1.4\\
		PKS 1502+106 & 226.10 & 10.50 & 9.8 &2.5 & 55509.5 & 21.6 & 1.97 & 1.8\\ 
		OT 081 & 267.87 & 9.65 & 9.7 &2.9 & 57751.6 & 45.7 & 0.79 & 1.3\\
		RX J1931.1+0937 & 292.78 & 9.63 & -- & -- & -- & -- & -- & --\\
		OG +050 & 83.18 & 7.55 & -- & -- & -- & -- & -- & --\\
		MGRO J1908+06 & 287.17 & 6.18 & 2.9 &2.1 & 57045.2 & 4.6 & 0.63 & 0.9\\
		PKS 0019+058 & 5.64 & 6.14 & -- & -- & -- & -- & -- & --\\
		\multirow{2}{*}{\textbf{TXS 0506+056}} & \multirow{2}{*}{\textbf{77.35}} & \multirow{2}{*}{\textbf{5.70}} &$\mathbf{10.0^{+5.2}_{-4.2}}$ & $\mathbf{2.2^{+0.3}_{-0.3}}$ & $\mathbf{57000^{+30}_{-30}}$ & $\mathbf{62^{+27}_{-27}}$ & \multirow{2}{*}{\textbf{2.77}} & \multirow{2}{*}{\textbf{1.7}}\\ & & & $\mathbf{7.6^{+6.1}_{-5.8}}$ & $\mathbf{2.6^{+0.5}_{-0.6}}$ & $\mathbf{58020^{+40}_{-40}}$ & $\mathbf{42^{+42}_{-28}}$ & & \\
		PKS 0502+049 & 76.34 & 5.00 & 2.7 &2.0 & 57072.1 & 1.2 & 0.81 & 0.9\\
		MG1 J123931+0443 & 189.89 & 4.73 & -- & -- & -- & -- & -- & --\\
		PKS 0829+046 & 127.97 & 4.49 & -- & -- & -- & -- & -- & --\\
		PKS 1502+036 & 226.26 & 3.44 & 2.0 &2.9 & 54606.9 & $3.4\times10^{-1}$ & 0.53 & 1.2\\
		HESS J1857+026 & 284.30 & 2.67 & 3.6 &2.3 & 54984.4 & $2.0\times10^{-1}$ & 0.71 & 0.9\\
		3C 273 & 187.27 & 2.04 & -- & -- & -- & -- & -- & --\\
		OJ 014 & 122.87 & 1.78 & -- & -- & -- & -- & -- & --\\
		PKS 0215+015 & 34.46 & 1.74 & -- & -- & -- & -- & -- & --\\
		PKS 0736+01 & 114.82 & 1.62 & -- & -- & -- & -- & -- & --\\
		4C +01.02 & 17.16 & 1.59 & -- & -- & -- & -- & -- & --\\
		4C +01.28 & 164.61 & 1.56 & -- & -- & -- & -- & -- & --\\
		GRS 1285.0 & 283.15 & 0.69 & 6.5 &2.8 & 54808.6 & 87.3 & 0.39 & 1.9\\
		PKS 0422+00 & 66.19 & 0.60 & -- & -- & -- & -- & -- & --\\
		PKS B1130+008 & 173.20 & 0.58 & -- & -- & -- & -- & -- & --\\
		PMN J0948+0022 & 147.24 & 0.37 & 2.0 &2.4 & 55610.7 & $4.3\times10^{-4}$ & 0.90 & 0.6\\ 
		HESS J1852-000 & 283.00 & 0.00 & 5.4 &2.8 & 54751.9 & 100.3 & 0.38 & 1.9\\
		\textbf{NGC 1068} & \textbf{40.67} & \textbf{-0.01} & $\mathbf{23.0^{+8.7}_{-7.9}}$ &$\mathbf{2.8^{+0.3}_{-0.3}}$ & $\mathbf{56290^{+90}_{-80}}$ & $\mathbf{198^{+64}_{-64}}$ & \textbf{2.65} & \textbf{1.9}\\
		HESS J1849-000 & 282.26 & -0.02 & -- & -- & -- & -- & -- & --\\
		PKS 0440-00 & 70.66 & -0.29 & 6.2 &2.6 & 57896.8 & 66.8 & 0.51 & 0.9\\
		PKS 1216-010 & 184.64 & -1.33 & -- & -- & -- & -- & -- & --\\
		PKS 0420-01 & 65.83 & -1.33 & -- & -- & -- & -- & -- & --\\
		NVSS J190836-012 & 287.20 & -1.53 & -- & -- & -- & -- & -- & --\\
		PKS 0336-01 & 54.88 & -1.77 & -- & -- & -- & -- & -- & --\\
		S3 0458-02 & 75.30 & -1.97 & 4.6 &2.5 & 56974.6 & $7.0\times10^{-1}$ & 0.65 & 0.7\\ 
		NVSS J141826-023 & 214.61 & -2.56 & 3.7 &2.9 & 57733.0 & $3.4\times10^{-1}$ & 0.44 & 0.6\\
		PKS 2320-035 & 350.88 & -3.29 & 10.8 &3.2 & 56176.8 & 160.2 & 0.57 & 1.1\\
		HESS J1843-033 & 280.75 & -3.30 & -- & -- & -- & -- & -- & --\\[3pt]
		\midrule
		PKS 1329-049 & 203.02 & -5.16 & -- & -- & -- & -- & -- & --\\
		HESS J1841-055 & 280.23 & -5.55 & -- & -- & -- & -- & -- & --\\
		3C 279 & 194.04 & -5.79 & -- & -- & -- & -- & -- & --\\
		HESS J1837-069 & 279.43 & -6.93 & -- & -- & -- & -- & -- & --\\
		PKS 0805-07 & 122.07 & -7.86 & -- & -- & -- & -- & -- & --\\
		PKS 1510-089 & 228.21 & -9.10 & -- & -- & -- & -- & -- & --\\
		PKS 0048-09 & 12.68 & -9.49 & -- & -- & -- & -- & -- & --\\
		PKS 0727-11 & 112.58 & -11.69 & -- & -- & -- & -- & -- & --\\
		PKS 2233-148 & 339.14 & -14.56 & 2.0 &2.8 & 54877.5 & $2.6\times10^{-3}$ & 1.04 & 12.0\\ 
		NGC 253 & 11.90 & -25.29 & 4.1 &2.5 & 56511.7 & 22.7 & 0.52 & 8.7\\
		NGC 4945 & 196.36 & -49.47 & 2.0 &1.9 & 54739.8 & $2.4\times10^{-1}$ & 0.63 & 55.3\\
		LMC & 80.00 & -68.75 & -- & -- & -- & -- & -- & --\\
		SMC & 14.50 & -72.75 & -- & -- & -- & -- & -- & --\\
		\hline\hline
		\caption{Coordinates (Right Ascension R.A. and declination $\delta$), maximum-likelihood flare parameters, logarithm of the local pre-trial p-values $p_{loc}$ of the sources of the catalog and the 90\% CL upper limits on the time-integrated flux $F_{90\%}$ (in units of TeV cm$^{-2}$) defined in equation~\ref{eq:time-int_flux_upLims} for an $E^{-2}$ spectrum. Under-fluctuating results are shown with hyphens. For the four sources that give rise to the $3.0~\sigma$ excess of the binomial test in the Northern hemisphere (highlighted in bold), the fit parameters are shown with the confidence interval at $68\%$ CL. A line is used to separate the Northern from Southern sources. The parameters of the flare from TXS 0506+056 at 58020 MJD and related to the neutrino alert ($n_s=7.6$, $\gamma=2.6$, $\sigma_T=42$ days) are different from those reported in \cite{IceCube:2018cha}, when the data available for analysis extended up to 40 days after the central time of the flare. This analysis includes 7 additional months and reconstructs a longer, more significant flare associated with the same alert.}
		\label{tab:PS_results1}
	\end{longtable}
\end{center}

\section{Multi-flare algorithm}
\label{sec:multi-flare_algorithm}


The multi-flare algorithm aims at determining the number of flares to fit in the data. This is done by evaluating the TS of clusters of events with the highest signal-over-background ratio of the spatial and energy PDFs and selecting as candidate flares those that pass a given TS threshold. On the one hand, a high value of TS threshold is required to suppress background fluctuations (fake flares), on the other hand a low value is desired to avoid the rejection of signal flares of low intensity.

\begin{figure}[ht]
	\centering
	\includegraphics[width=.80\linewidth]{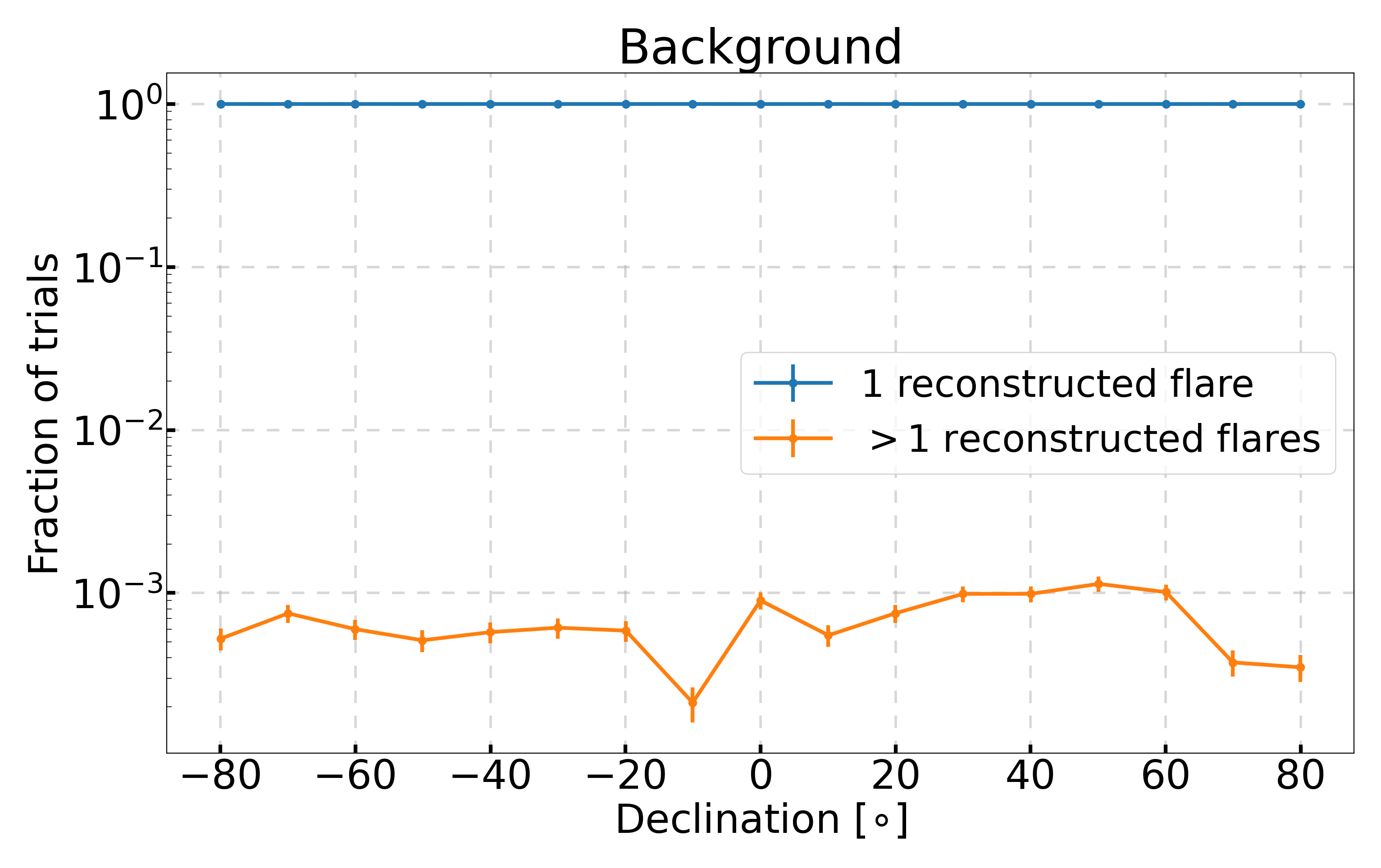} 
	\caption{Fraction of trials in which, under the null hypothesis, 1 single flare (blue line) or more than 1 flare (orange line) are reconstructed as a function of the declination if a TS threshold of 2 is applied to select the candidate flares.}
	\label{fig:bkg_reco_flares}
\end{figure}

\begin{figure}[ht]
	\centering
	\includegraphics[width=.49\linewidth]{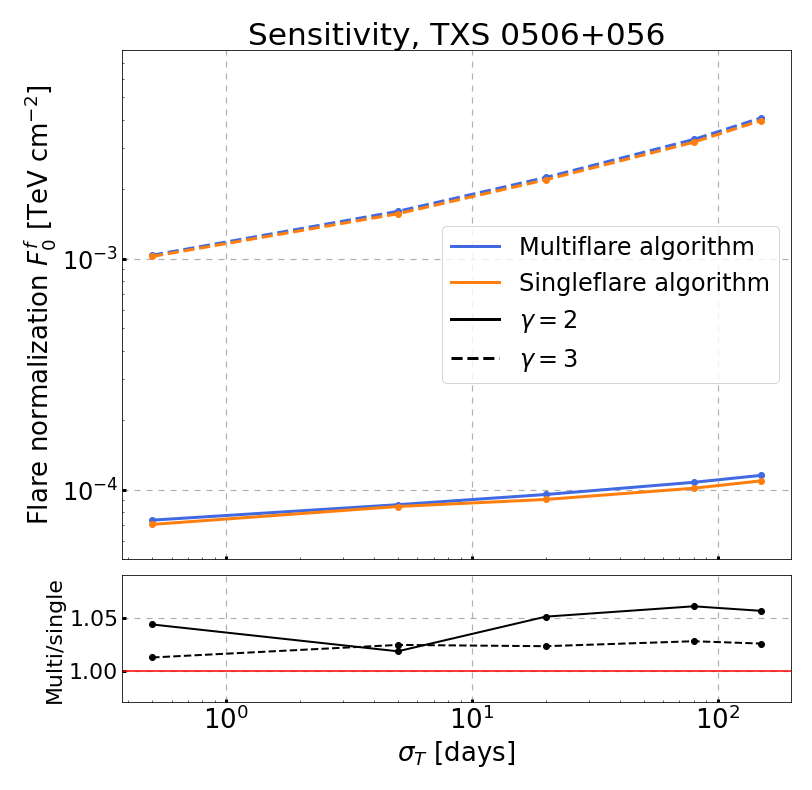}
	\includegraphics[width=.49\linewidth]{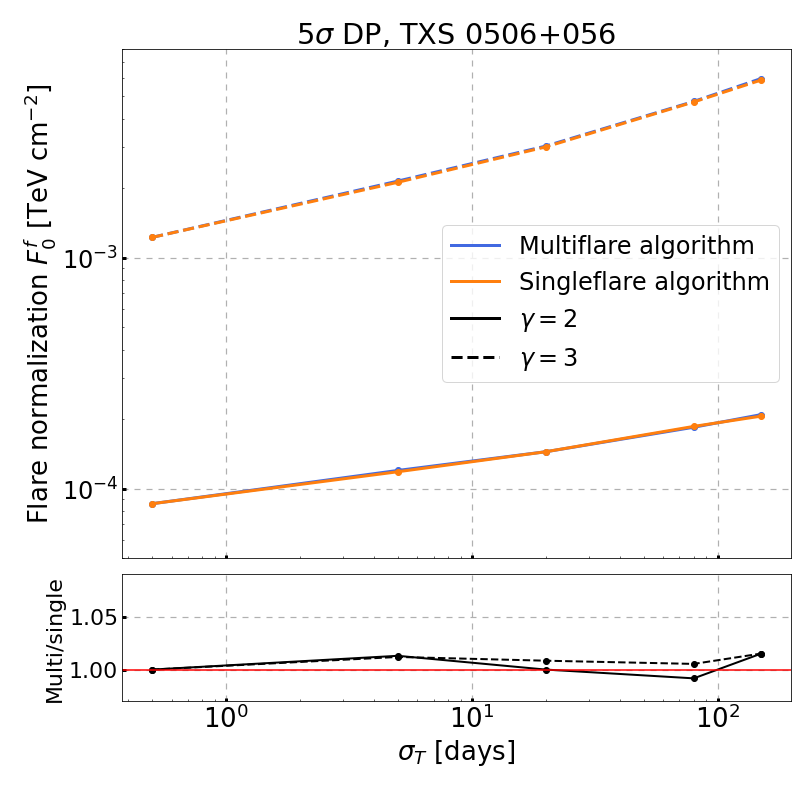}
	\caption{Sensitivity (left) and discovery potential (right) of the single-flare (orange lines) and multi-flare (blue lines) algorithm as a function of the flare duration $\sigma_T$. Sensitivity and discovery potential are evaluate at the declination of TXS 0506+056 under the hypothesis of a single signal flare with a spectrum $E^{-2}$ (solid lines) and $E^{-3}$ (dashed lines). The bottom plots show the ratio of the multi-flare to single-flare curves above.}
	\label{fig:single_VS_multi_sensDP}
\end{figure}

This multi-flare algorithm selects as candidate flares the cluster of events with the highest TS and all additional clusters of events passing a TS threshold of 2. The choice of this threshold ensures a high efficiency in rejecting fake flares, with a frequency of multiple flare reconstruction under the null hypothesis of $\lesssim 0.1\%$ as shown in Fig.~\ref{fig:bkg_reco_flares}. Such a high rejection efficiency allows to preserve a sensitivity and a DP comparable to the single-flare algorithm, as shown in Fig.~\ref{fig:single_VS_multi_sensDP} at the declination of TXS 0506+056. Note additionally that if only one candidate flare is selected by the multi-flare algorithm, the multi-flare algorithm is completely equivalent to the single-flare algorithm. 



\begin{figure}[ht]
	\centering
	\includegraphics[width=.49\linewidth]{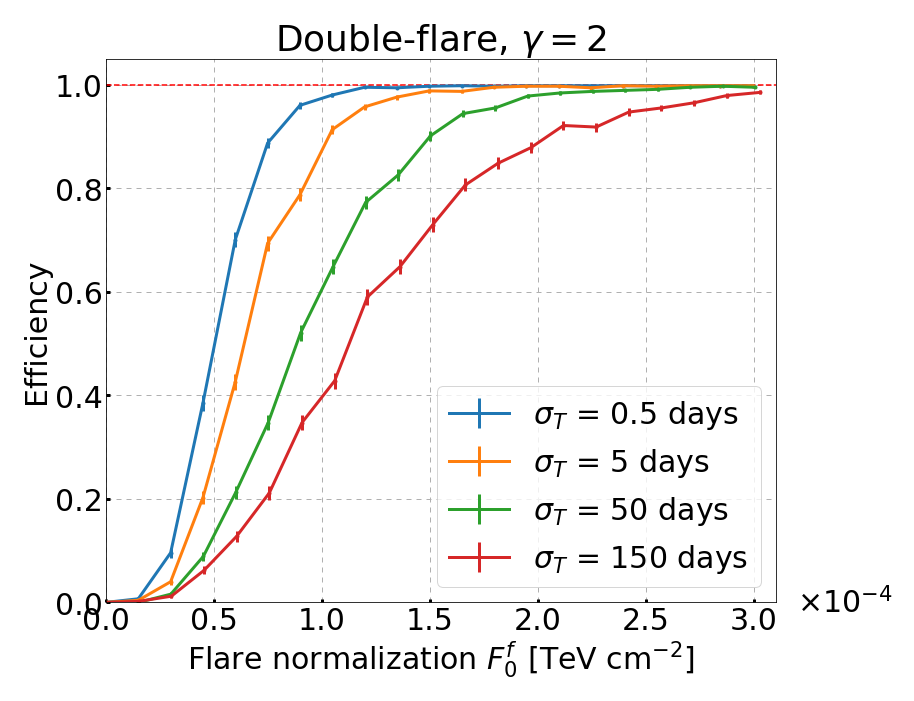} 
	\includegraphics[width=.49\linewidth]{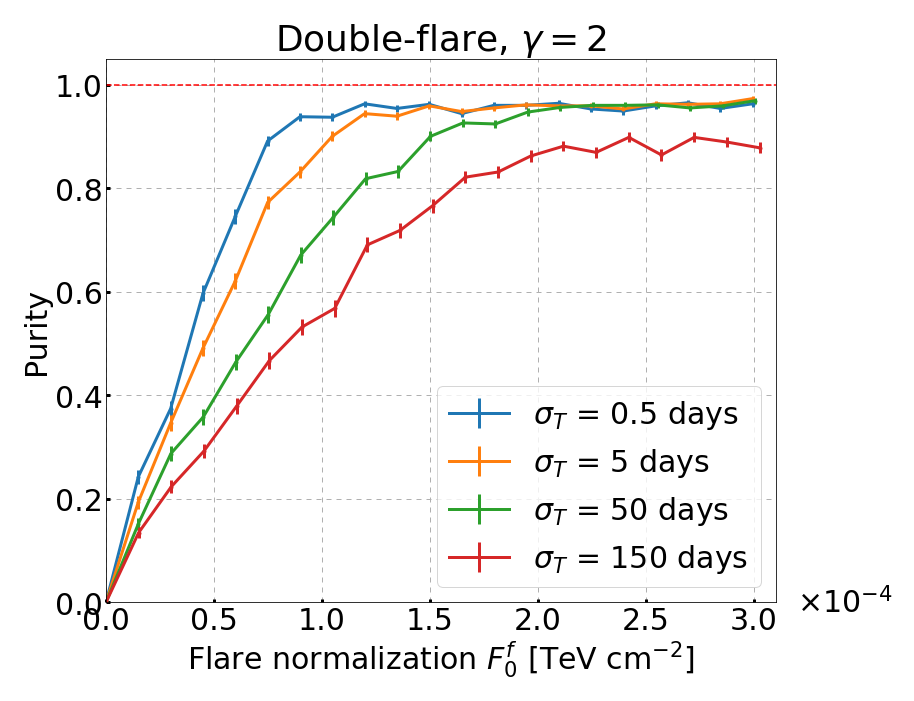}
	\includegraphics[width=.49\linewidth]{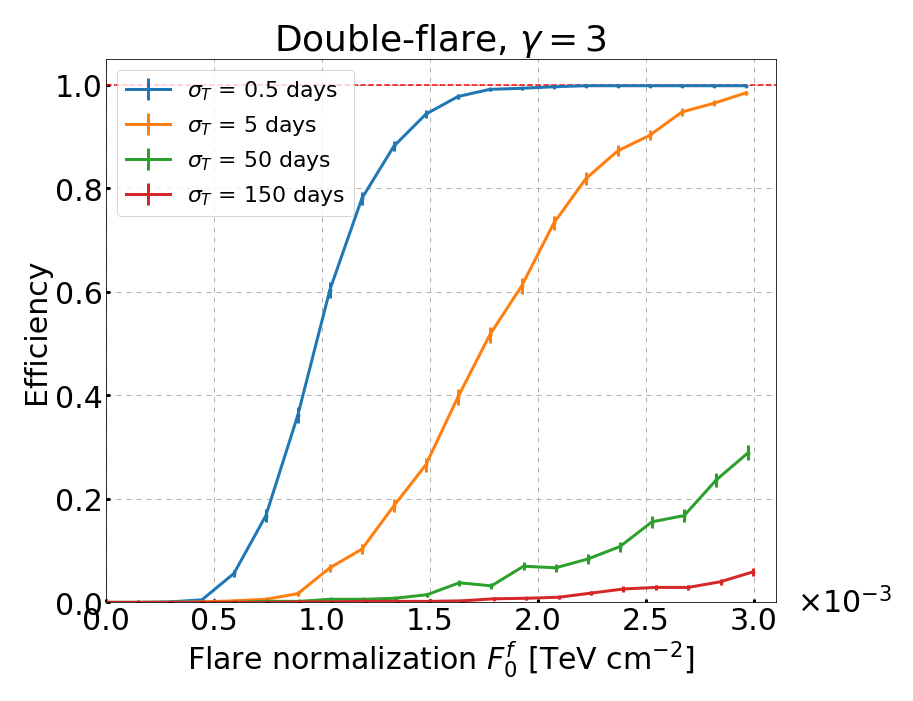} 
	\includegraphics[width=.49\linewidth]{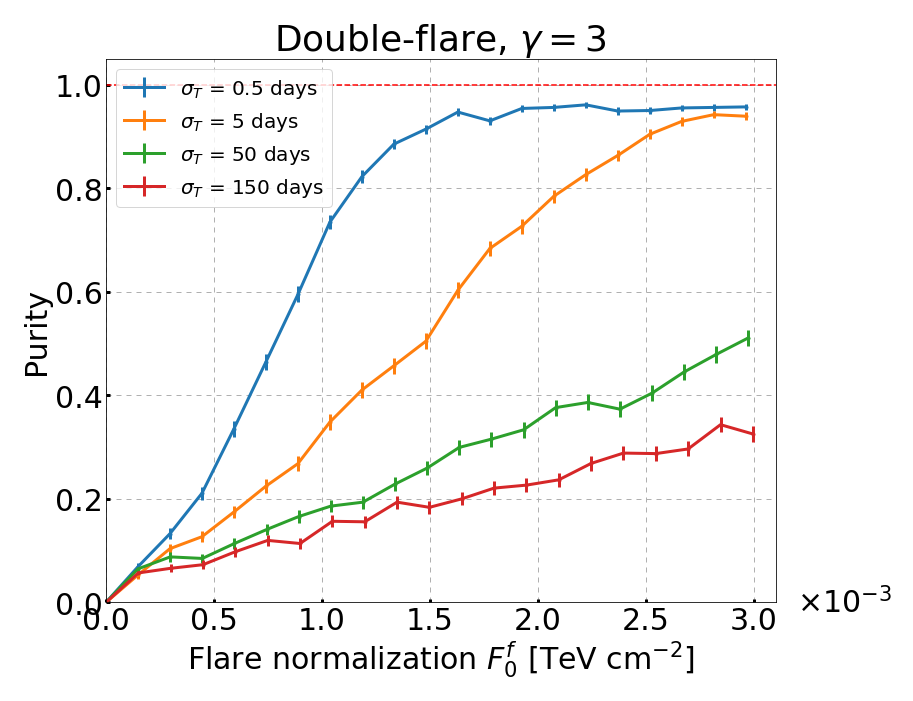}
	\caption{Efficiency (left plots) and purity (right plots) under the hypothesis of a two flares of equal intensity as a function of the time-integrated flux of each flare. Efficiency and purity are calculated for a spectrum $E^{-2}$ (top plots, declination of TXS 0506+056) and $E^{-3}$ (bottom plots, declination of NGC 1068) and for some values of $\sigma_T$ (see legend). Efficiency is defined as the fraction of trials in which \textit{all} the injected flares are correctly reconstructed (no matter if additional fake flares are also reconstructed); purity is defined as the fraction of trials in which no fake flares are reconstructed. Note that a partially reconstructed flare is considered as a fake flare when calculating the efficiency and purity.}
	\label{fig:efficiency_and_purity}
\end{figure}

To quantify the goodness of the multi-flare reconstruction, two quantities are introduced: the multi-flare efficiency, defined as the fraction of trials in which all the signal flares are identified (no matter if additional fake flares are also reconstructed), and the multi-flare purity, defined as the fraction of trials in which no fake flares are reconstructed (no matter if all the signal flares are identified). The former is an indicator of how frequent the algorithm is able to identify \textit{all} the signal flares injected in the data; the latter is an indicator of how well the algorithm is able to reject fake flares. Note that a partially reconstructed flare is considered as a fake flare in the estimation of efficiency and purity. These two quantities are shown in Fig.~\ref{fig:efficiency_and_purity} under the hypothesis of two flares of equal intensity as a function of the time-integrated flux of each flare, for spectra $E^{-2}$ and $E^{-3}$ and for some values of $\sigma_T$. The efficiency asymptotically reaches the value of 1: if the signal is strong enough the algorithm is always able to identify it. However, the flux required to reach such asymptotic \textit{plateau} depends on the parameters of the flare (spectral index $\gamma$ and flare duration $\sigma_T$), and notably in extreme cases (soft spectra, long flare duration) the convergence is very slow, as a consequence of the high TS threshold. Nevertheless, note that in such extreme cases the flare intensity is mostly below the sensitivity level. The purity also tends to an asymptotic \textit{plateau} at $\gtrsim 95\%$ with a rapidity that depends on the flare parameters.

\section{\textit{A posteriori} comparisons with the time-integrated analysis}
\label{app:variab}

The results of these time-dependent analyses, despite unveiling new features of the source catalog, partly overlap with the results of the time-integrated search~\citep{Aartsen:2019fau}. In fact, the time-dependent and time-integrated analyses are based on similar likelihood functions, sharing the same space and energy PDFs, but the time-dependent analysis distinguishes itself by adding a time PDF. This time-dependent analysis was planned together with the time-integrated analysis, and it was not developed based on the time-integrated unblinded results. Nonetheless, one might wonder how the results of the time-dependent analysis can be interpreted in the light of the prior knowledge of the time-integrated results. To address such a question, two tests are proposed in this Appendix. A first test estimates the time variability of the four most significant sources of the time-integrated analysis. A second test estimates the probability of obtaining the observed pre-trial significance of $3.8~\sigma$ from a time-dependent binomial test (see Section~\ref{sec:results}) on the source catalog, in the assumption that the neutrino excess observed by the time-integrated analysis~\citep{Aartsen:2019fau} does not have any time structure. Both tests exploit the same approach, based on producing pseudo-realizations of the data by randomizing the time of the events and, unlike for the standard time-dependent analysis, keeping fixed the associated equatorial coordinates.

\paragraph{\bf Time-variability test:}
This test aims at quantifying the time variability of the highly-significant events detected from the directions of NGC 1068, TXS 0506+056, PKS 1424+240 and GB6 J1542+6129 and at testing the compatibility of their arrival time with a flat distribution.
This test is sensitive only to the time information of the events and is unavoidably less sensitive than the time-dependent search described in Section~\ref{sec:analysis} (referred to as standard time-dependent analysis), which is sensitive to energy, space and time information. Moreover, the significance of the likelihood method using the three variables at the same time is not equivalent to the product of likelihood methods that use one variable at a time. 

The null (or background) hypothesis of the time-variability test assumes that the time-integrated signal-like events (i.e. the events with the highest time-integrated signal-over-background ratio, that mostly contribute to the significance around each source direction) are not clustered in time. Pseudo-realizations of the data for this null hypothesis (also called background samples which allow to count trials) are obtained similarly to the standard time-dependent analysis: events in a declination band around the location of the tested sources are selected and assigned a new time taken from a real up time of the detector. This procedure destroys any time correlation among events. However, while the standard analysis keeps the local coordinates of an event (azimuth and zenith) fixed and recalculates the right ascension using the new randomized time, the time-variability test freezes the equatorial coordinates of the events at the measured values, and randomizes the azimuth (notably the zenith angle, corresponding to an equatorial coordinate, at the South Pole does not depend on the time). This method guarantees that the same time-integrated signal-like events from the direction of a given source are present in the background sample with randomized times. On the other hand, this method flattens out the sub-daily modulation of the event rate in local coordinates due to the increased reconstruction efficiency along azimuth directions where more strings are aligned. As described in Section~\ref{sec:analysis}, in the standard analysis this sub-daily modulation of the event rate is taken into account by using a correction in local coordinates to the background PDF. The azimuth dependency of the reconstruction efficiency is averaged out for flares longer than $\sigma_T=0.2$ days as a consequence of the Earth rotation, while it might induce a change up to 5\% in the TS for flares as short as $\sigma_T = 10^{-3}$ days. Given that the variability observed for the four most significant time-integrated sources was beyond a flare duration of $\sigma_T\gg 1$ day, a lower limit $\sigma_T^{min} = 0.2$ days is used for this time-variability test.

Whereas for the standard analysis signal samples are produced by injecting Monte Carlo events on top of the background events, for the time-variability test $n_s$ events among real signal-like events are selected in the data and their times are sampled from a Gaussian distribution. The real signal-like events, potentially usable for signal injection in the time-variability test, are randomly chosen among the $2\hat{N}_s^{t-int}$ events with the highest time-integrated signal-over-background ratio, where $\hat{N}_s^{t-int}$ is the best-fit number of signal-like events reported by the time-integrated analysis~\citep{Aartsen:2019fau}.

\begin{figure}[htbp]
	\centering
	\includegraphics[width=.95\linewidth]{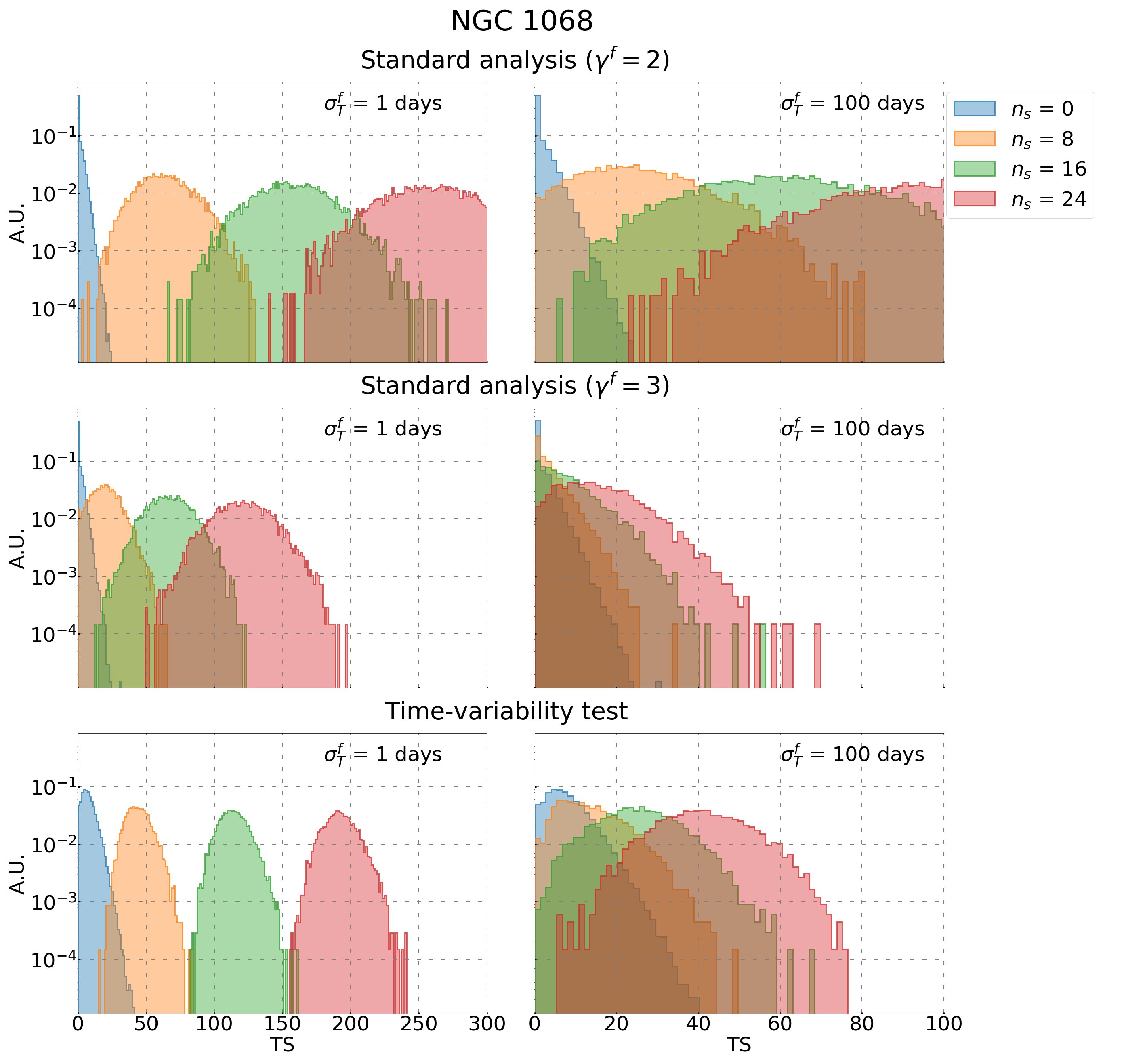}
	\caption{Comparison of the TS distributions for signals of different intensity $n_s$ and for the background between the standard analysis (first and second row) and the time-variability test (third row) at the location of NGC 1068. The left plots are made for a flare duration of $\sigma_T=1$~d, the right plots for 100~d. Spectral indices of $\gamma^f=2$ (first row) and $\gamma^f=3$ (second row) are used for the injected signal in the standard analysis.}
	\label{fig:ts_comparison}
\end{figure}

\begin{figure}[htbp]
	\centering
	\includegraphics[width=.6\linewidth]{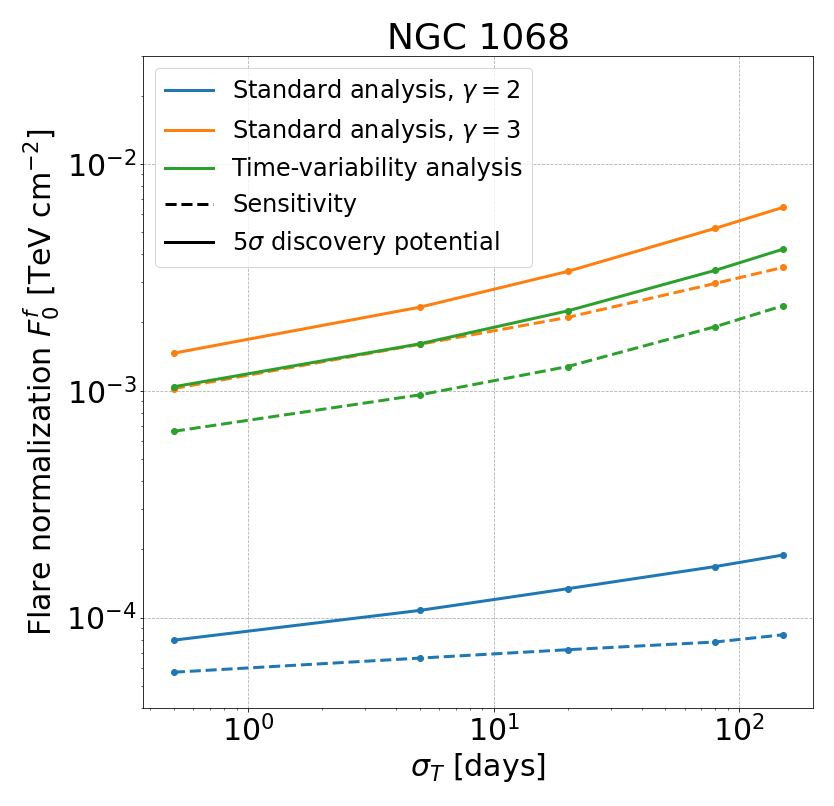}
	\caption{Comparison of the sensitivity (dashed lines) and $5~\sigma$ DP (solid lines) of the standard analysis (blue and orange, respectively for $\gamma=2$ and $\gamma=3$) and the time-variability test (green lines) in terms of the time-integrated flux per flare $F_0^f$ described in equation~\ref{eq:time-integrated_flux}. These curves are produced at the location of NGC 1068 under the hypothesis of a single signal flare. Notice that the reconstructed value of the spectral index for NGC 1068 in \cite{Aartsen:2019fau} is 3.16.}
	\label{fig:time-variability_comparison}
\end{figure}

The likelihood in Eq.~\ref{eq:10-year-likelihood} is maximized on the background and signal samples of the time-variability test and the corresponding TS distributions (for illustration at the location of NGC 1068) are shown in Fig.~\ref{fig:ts_comparison}, for comparison with the same distributions for the standard analysis. For both analyses the separation of the signal and background TS is better for shorter flares (left plots) than longer ones (right plots). A notable feature concerns the background TS distributions in blue. For the standard analysis the TS distribution has a characteristic spike in the first bin populated by under-fluctuations set to zero. On the other hand, the TS distribution for the time-variability test is on average shifted towards larger values of TS, showing a more signal-like behavior. This is a consequence of preserving the same time-integrated space and energy variables of signal-like events in the background sample with the method described above. 
It is to be noted that the time-integrated analysis in~\cite{Aartsen:2019fau} fits a spectral index of NGC 1068 of 3.16, while the best-fit spectral index for the time-dependent analysis is harder, namely 2.8 (see Tab.~\ref{tab:PS_results1}). As a consequence of preserving the spatial and energy information of the events, the background and signal samples of the time-variability test (used to make the distributions in the last row in Fig.~\ref{fig:time-variability_comparison}) have a varying spectral index centered around 2.8. Notably, about 89\% of the spectral indices of the 100,000 generated background samples are contained between $\gamma^f=2$ and of $\gamma^f=3$. Hence, these values of the spectral indices are used for the signal injection in the standard analysis when comparing the TS distributions of the standard analysis with the same distributions of the time-variability test in Fig.~\ref{fig:time-variability_comparison}. In general, for harder spectral indices and the same flare duration $\sigma_T^f$, the time-variability test characterizes the difference between background and signal less powerfully than the standard analysis. In fact, in the time-variability test the coordinates of the events are frozen to the true values, hence the differences between the spatial and energy PDFs of signal and background are not exploited, unlike for the standard analysis.

To complete the comparison between the standard time-dependent analysis and the time-variability test, the sensitivity and $5~\sigma$ DP at the location of NGC 1068 are shown for the two analyses in Fig.~\ref{fig:time-variability_comparison}. The times of signal events are sampled from a Gaussian distribution with fixed mean $t_0=58000$ MJD and variable width $\sigma_T$. This plot can be understood by observing the TS distributions in Figure~\ref{fig:ts_comparison}.

For each of the four aforementioned sources, the likelihood in Eq.~\ref{eq:10-year-likelihood} is maximized on the real data and an observed TS is reported. A pre-trial p-value for the time-variability test is then evaluated by counting the fraction of generated background samples with TS larger than the observed TS. The post-trial p-value of each source is obtained by applying a Sidak correction (\cite{Abdi2007}) to the pre-trial p-values with penalty factor 4 (the number of sources). The results of this test are shown in Table~\ref{tab:time_var_analysis}. None of the four sources shows a significant time variability for the signal-like neutrino events. 

\begin{table}
	\centering
	\begin{tabular}{>{\centering\arraybackslash}m{2.8cm} >{\centering\arraybackslash}m{2.8cm} >{\centering\arraybackslash}m{2.8cm} } 
		\multicolumn{3}{c}{Time-variability results}\\
		\hline
		\hline
		Source & Pre-trial p-value & post-trial p-value\\[3pt] \hline
		\vspace{3pt}
		NGC 1068 & 0.13 & 0.43 \\[3pt] 
		TXS 0506+056 & 0.24 & 0.67\\  [3pt] 
		PKS 1424+240 & 0.33 & 0.80 \\[3pt] 
		GB6 J1542+6129 & 0.029 & 0.11 \\[3pt] 
		\hline
		\hline
	\end{tabular}
	\caption{Results of the time-variability test applied to the four most significant sources of the time-integrated analysis of Ref.~\cite{Aartsen:2019fau}. The table shows the p-values before (pre-trial) and after (post-trial) the Sidak correction with penalty factor 4. As described in this Appendix, the time-variability test only assesses the time distribution of the recorded events, by comparing with simulated samples in which the event directions and energies remain fixed as recorded, but times are randomized according to a uniform distribution.}
	\label{tab:time_var_analysis}
\end{table}

It is worth noticing the case of M87: this source was an under-fluctuation for the time-integrated analysis, with no signal-like events identified in \citep{Aartsen:2019fau}, but it shows up as the most significant source of the catalog when a time-dependent analysis is performed. Although a time-variability test is not made in this case, with $\hat{n}_s=3$ signal-like neutrino events in a time scale of $\hat{\sigma}_T=2.0$ minutes, almost the entire significance of this source is expected to come from the time variability of the detected events.

\paragraph{\bf Posterior time-dependent binomial test:} The second test determines the \textit{a posteriori} probability that the time-dependent binomial test (see Section~\ref{sec:analysis} referred to as “standard” in this Appendix) produces a pre-trial significance as high or higher than the observed value of $3.8~\sigma$, in the assumption that the time-integrated neutrino excess is steady in time (background hypothesis). To do so, the same binomial test described in Section~\ref{sec:analysis} is repeated on the list of p-values of the Northern sources. The per-source p-values are computed in the same way, by comparing the TS of each source with a distribution of TS from fully-scrambled (randomized times and recalculated right ascensions) background samples at the respective declination. As a matter of fact, the binomial p-value of the data for this test (referred to as “posterior binomial test”) is the same as reported in Section~\ref{sec:results} ($3.8\sigma$). Nevertheless, the difference between the standard and the posterior binomial test is in the realization of the background samples used to translate the binomial p-value into a post-trial p-value.

In the posterior binomial test, background pseudo-realizations of the data for all the Northern sources of the catalog are obtained in the same way as described for the time-variability test: the times of the events are randomized while the equatorial coordinates are fixed at the recorded values, such that the spatial correlations among the events are preserved and the time-integrated information is effectively incorporated in the background hypothesis. For each pseudo-realization of the Northern catalog, the likelihood in Eq.~\ref{eq:10-year-likelihood} is maximized at the location of each source and the corresponding TS is converted into a pre-trial p-value as described in Section~\ref{sec:analysis}, by comparison with a distribution of TS from fully-scrambled background samples at the same declination. The lower limit on the flare duration $\sigma_T^{min}$ is removed in this test to allow a proper comparison with the standard time-dependent binomial test. As a consequence, the azimuth-dependent correction to the background spatial PDF is neglected. However, this is a minor correction that has an impact at most of 5\% only for time scales of the flares as short as $\sim10^{-3}$ days.

\begin{figure}[htbp]
	\centering
	\includegraphics[width=.8\linewidth]{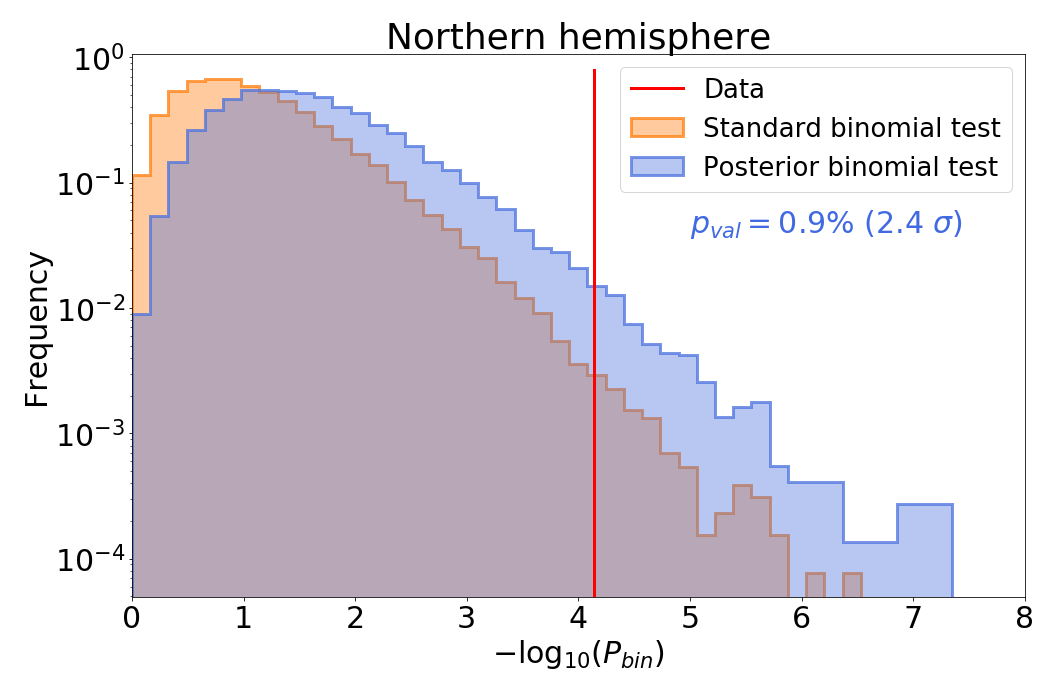}
	\caption{Background distribution of the binomial p-value $P_{bin}$ for the posterior (blue) and standard (orange) binomial test. For the posterior binomial test, the background sample is produced by randomizing the time of the events while keeping fixed the equatorial coordinates; for the standard analysis, the equatorial coordinates are recalculated (assuming fixed local coordinates) after the time is randomized.}
	\label{fig:binomial_p-value_distr}
\end{figure}

Once a pre-trial p-value is computed for all the sources in a particular pseudo-realization of the Northern catalog, the binomial test is performed on this set of p-values, resulting in a background binomial p-value $P_{bin}$. This method is then repeated on many pseudo-realizations of the Northern catalog to produce the distribution of background binomial p-values for the posterior binomial test shown in blue in Fig.~\ref{fig:binomial_p-value_distr}. These p-values are the typical binomial p-values that the binomial test produces if the neutrino events of the data (including the time-integrated excess) have no time structure. For comparison, the orange histogram in Fig.~\ref{fig:binomial_p-value_distr} is the distribution of background binomial p-values for the standard binomial test, used in Section~\ref{sec:results} to calculate the post-trial binomial p-value in the assumption that the time-integrated information is also randomized. Note that when the time-integrated information is preserved, the overall distribution is shifted towards higher values of $-\log_{10}(P_{bin})$ as a consequence of including the additional information about the time-integrated excess in the background samples.

Finally, the probability that the time-dependent binomial test produces a more significant result than the one observed in the real data ($3.8~\sigma$ pre-trial), given the prior knowledge about the time-integrated excess and under the assumption that the neutrino events do not have any time correlation, is estimated by counting the fraction of background binomial p-values of the posterior binomial test that are more significant than the observed result. Such estimation leads to a probability of $0.9\%$.

\section{Estimation of the single-flare significance of TXS 0506+056}
\label{sec:singleflare_significance}

This Appendix is intended to describe how the single-flare significances of the two flares of TXS 0506+056, that are shown in Fig.~\ref{fig:best_fit_flares}, are estimated.

By factorizing the background PDF, the multi-flare likelihood ratio $\Lambda_{mf}^{-1}$ in Eq.~\ref{eq:teststatistic} can be written as follows:
\begin{equation}
    \label{eq:likelihood_ratio_simple}
    \Lambda^{-1}_{mf}=\frac{\mathcal{L}(\vec{\hat{n}}_s, \vec{\hat{\gamma}}, \vec{\hat{t}}_0,\vec{\hat{\sigma}}_T)}{\mathcal{L}(n_s=0)}=\prod_{j=\mathrm{sample}}\prod_{i=\mathrm{1}}^{N_j}\left(1+\sum_{f=\mathrm{flares}}\mathcal{F}^f_{i,j}\right), \ \ \ \ \ \ \ \ \ \ \ \ \ \ \ \ \ \ \ \ \mathcal{F}_{i,j}^f \coloneqq\frac{n_s^f(\mathcal{S}^f_{i,j}/\mathcal{B}_{i,j}-1)}{N}.
\end{equation}
The single-flare signal and background PDFs in Eq.~\ref{eq:likelihood_ratio_simple} are the same as in Eq.~\ref{eq:multi-likelihood}, but for the sake of clarity here they explicitly show the flare ($f$), event ($i$) and sample ($j$) indices. In addition, the dependency on the parameters, being the same as in Eq.~\ref{eq:multi-likelihood}, is omitted to simplify the notation.

For TXS 0506+056 there are two identified flares, thus $\sum_f \mathcal{F}^f_{i,j}=\mathcal{F}^1_{i,j}+\mathcal{F}^2_{i,j}$. In addition, when an event $i$ does not contribute significantly to $\mathcal{F}^f_{i,j}$, then $\mathcal{F}^f_{i,j}\sim10^{-6}\text{--}10^{-4}$. Since an event can contribute significantly only to one flare, the crossed terms $\mathcal{F}^1_{i,j}\mathcal{F}^2_{i,j}$ can be neglected and it is meaningful to retain only terms at first order in $\mathcal{F}^f_{i,j}$. Based on these observations, the likelihood ratio in Eq. \ref{eq:likelihood_ratio_simple} can be well approximated as:
\begin{equation}
    \Lambda^{-1}_{mf}=\prod_{j=\mathrm{sample}}\prod_{i=\mathrm{1}}^{N_j}\left(1+\mathcal{F}^1_{i,j}+\mathcal{F}^2_{i,j}\right)\simeq
   \prod_{j=\mathrm{sample}}\prod_{i=\mathrm{1}}^{N_j}\left(1+\mathcal{F}^1_{i,j}\right)\left(1+\mathcal{F}^2_{i,j}\right)=\left(\Lambda_{sf}^{f=1}\right)^{-1}\left(\Lambda_{sf}^{f=2}\right)^{-1}.
\end{equation}
Thus, it can be factorized into single-flare components, that are equivalent to the multi-flare likelihood ratio when only one flare is considered. This result can be easily generalised to $N_f>2$ flares.

Such a factorization can be exploited to disentangle the contribution of each flare to the multi-flare TS in Eq. \ref{eq:teststatistic}:
\begin{equation}
    \mathrm{TS}\simeq-2\log\left[\frac{1}{2}\prod_{f=\mathrm{flares}}\left(\frac{T_{live}}{\hat{\sigma}_T^fI\left[\hat{t}_0^f,\hat{\sigma}_T^f\right]}(\Lambda_{sf}^f)^{-1}\right)\right]=-2\sum_{f=\mathrm{flares}}\log\left[\left(\frac{1}{2}\right)^{1/N_f}\frac{T_{live}}{\hat{\sigma}_T^fI\left[\hat{t}_0^f,\hat{\sigma}_T^f\right]}(\Lambda_{sf}^f)^{-1}\right]=\sum_{f=\mathrm{flares}}\mathrm{TS}_{sf}^{f},
\end{equation}
where $\mathrm{TS}_{sf}^f$ is the contribution of the $f$-th flare to the multi-flare TS and can be interpreted as a single-flare TS.

The single-flare significance $\sigma_{sf}^f$ can be obtained in the same way as the multi-flare significance, but using the single-flare TS instead of the multi-flare TS. Assuming that the two flares of TXS 0506+056 are independent, one might expect to retrieve the multi-flare TS by summing linearly the single-flare TS and to retrieve the multi-flare significance $\sigma_{mf}$ by summing in quadrature the single-flare significance. Although this is effectively observed for the TS, the summation in quadrature of the single-fare significance results in a mismatch of nearly 2.5\% with respect to the multi-flare significance. To correct for this mismatch, the single-flare significance is redefined as $\sigma^{\prime f}_{sf}$ through the following relation:
\begin{equation}
    \frac{\sigma^{\prime 1}_{sf}}{\sigma^{\prime 2}_{sf}}=\frac{\sigma^1_{sf}}{\sigma_{sf}^2}, \ \ \ \ \ \ \ \ \ \ \ \ \ \ \ \ \sqrt{\left(\sigma^{\prime 1}_{sf}\right)^2+\left(\sigma^{\prime 2}_{sf}\right)^2}=\sigma_{mf}.
\end{equation}

For TXS 0506+056 this method is used to disentangle the single-flare significance $\sigma^{\prime f}_{sf}$ of the 2 flares used in Fig.~\ref{fig:best_fit_flares}.

\section{Investigation of the significance of TXS 0506+056}
\label{sec:TXS_significance_investigation}
The significance of TXS 0506+056 found by this multi-flare algorithm is smaller than the (single-flare) time-dependent significance that was determined in \cite{IceCube:2018cha}. The goal of this Appendix is to show that the decrease of significance is only due to the different event selection of the sample used in this analysis, and not due to the different likelihood algorithms. It is mainly related to 2 cascade events that are rejected in the new event selection, presented in~\citep{Aartsen:2019fau}. This was discussed also in IceCube~\citep{Abbasi:2021bvk}. As a matter of fact, the new selection was focused on muon tracks for achieving best angular resolutions for the point-source search.

The differences between this analysis and the one described in \cite{IceCube:2018cha} are mainly of three types. These are investigated using the analysis described in this letter and the one presented in \cite{IceCube:2018cha} to find out how much each of them contributes to the change in significance of TXS 0506+056. The results are summarized in Table~\ref{tab:TXS_comparisons}.

\paragraph{\textbf{Different datasets:}}
    As mentioned also in Section \ref{sec:detector}, the event selections used to produce the dataset analyzed in \cite{IceCube:2018cha} and the one analyzed in this work (from~\cite{Aartsen:2019fau}) are different. According to the internal IceCube nomenclature, the two datasets are referred to as \MA{{\tt PSTracks v2}} and \MA{{\tt PSTracks v3}}, respectively. In some cases the different event selection results in the reconstruction of slightly different energy and local angles. An extensive and detailed description of the two datasets can be found in~\cite{Abbasi:2021bvk}.
    
     The significance of TXS 0506+056 is estimated on \MA{{\tt PSTracks v2}} and \MA{{\tt PSTracks v3}} by applying the multi-flare algorithm to the years 2012-2015 (containing only one of the two flares detected by this analysis). We observe the same drop in significance (from $4.0~\sigma$ in \MA{{\tt PSTracks v2}} to $2.6~\sigma$ in \MA{{\tt PSTracks v3}}) described in~\cite{Abbasi:2021bvk}. The significance observed for \MA{{\tt PSTracks v3}} increases to $3.4~\sigma$ if the two high-energy events, present in \MA{{\tt PSTracks v2}} but absent in \MA{{\tt PSTracks v3}}, are added by hand to the dataset. It is worth noticing also that the pre-trial significance observed for \MA{{\tt PSTracks v2}} with the multi-flare algorithm is not different to the pre-trial significance reported in \citep{IceCube:2018cha}, which was obtained with a single-flare algorithm.

    \paragraph{\textbf{Different algorithms:}}
    The multi-flare algorithm has been developed for this analysis and applied for the first time in this work. 
    This is a crucial difference between this work and the one presented in \cite{IceCube:2018cha}, since a multi-flare likelihood could in principle consist of more fit parameters than a single-flare likelihood. The increased parameter space of the fit may thus degrade the sensitivity. This degradation was avoided by requiring a pre-selection of candidate flares with $\mathrm{TS}\ge2$ (see Section \ref{sec:analysis} and Appendix~\ref{sec:multi-flare_algorithm}).
    Other minor improvements between the two analyses concern:
    a Gaussian integral factor, included in the marginalization term to correct for boundary effects;
    the time PDF normalization, set to 1 across each IceCube sample by considering only up times of the detector (in \cite{IceCube:2018cha} it was set to 1 in an infinite range, regardless the up times). The results, shown in Table~\ref{tab:TXS_comparisons} for the single- and multi-flare algorithm applied to the 2012-2015 data, suggest that the multi-flare algorithm is not responsible for the drop of the significance, when applied to the same dataset. 

    \paragraph{\textbf{Different strategies for combining independent samples:}} 
    The third and last potential source of change in significance is due to the different strategies adopted to combine the IceCube samples.
    Since the  10-year data sample of IceCube concerns different IceCube detector configurations, triggers and event cuts, this analysis is based on the maximization of the joint likelihood defined as the product of the likelihoods of each IceCube sample (see Section \ref{sec:analysis}). The strategy adopted in~\cite{IceCube:2018cha}, instead, consists in maximizing the likelihood of each IceCube sample, picking up the most significant p-value and reporting it as post-trial after correcting for the look-elsewhere effect. Such a correction is made by penalizing the most significant $p$-value by the ratio of the livetime of the sample with the most significant $p$-value to the total time. To investigate this difference, the single-flare algorithm is applied to \MA{{\tt PSTracks v3}}. To reproduce the analysis in~\citep{IceCube:2018cha}, the TS is maximized only across the 3 years between 2012-2015 (containing the most significant flare) and the $p$-value is penalized by the ratio of 10 years to 3 years, adopting the same logic described in \cite{IceCube:2018cha}. In the analysis presented in this letter, the whole 10-year data are analyzed with a single joint likelihood (as described in Section \ref{sec:analysis} but without the multiple flare feature), and the same penalization of the $p$-value is not needed in this case. As seen in Table~\ref{tab:TXS_comparisons}, it can be stated that the results obtained in the two cases are comparable and that the strategy adopted to combine the different samples is not responsible for a substantial change in significance.

\begin{table}[h]
\centering
\begin{tabular}{>{\centering\arraybackslash}m{5cm} >{\centering\arraybackslash}m{3.5cm} >{\centering\arraybackslash}m{3.5cm}}
    \multicolumn{3}{c}{TXS 0506+056 change in significance}\\
    \hline
    \hline
    \multirow{3}{*}{\parbox{4.2cm}{\centering Different datasets (multi-flare, 2012-2015 only)}} & \multirow{2}{*}{\parbox{3.5cm}{\centering \MA{{\tt PSTracks v2}}\\(\cite{IceCube:2018cha})}} & \multirow{2}{*}{\parbox{3.5cm}{\centering \MA{{\tt PSTracks v3}}\\(This work)}}\\
    & & \\
    & $4.0~\sigma$ & $2.6~\sigma$ \\[3pt]
    \hline
    \multirow{6}{*}{\parbox{3cm}{\centering Different algorithms (2012-2015 only)}} &\multirow{2}{*}{\parbox{3.5cm}{\centering Single-flare\\(\cite{IceCube:2018cha})}} & \multirow{2}{*}{\parbox{3.5cm}{\centering Multi-flare\\(This work)}}\\
    & & \\
    & \multicolumn{2}{c}{\MA{{\tt PSTracks v2}}}\\
    & $4.0~\sigma$ & $4.0~\sigma$ \\
    & \multicolumn{2}{c}{\MA{{\tt PSTracks v3}}}\\
    & $2.7~\sigma$ & $2.6~\sigma$ \\
    \hline
    \multirow{3}{*}{\parbox{5cm}{\centering Strategy of sample combination (single-flare, \MA{{\tt PSTracks v3}})}} & \multirow{2}{*}{\parbox{3.5cm}{Separate likelihoods\\(\centering\cite{IceCube:2018cha})}} & \multirow{2}{*}{\parbox{3.5cm}{\centering Joint likelihood\\(This work)}}\\
    & & \\
    & $2.2~\sigma$ (post-trial) & $2.3~\sigma$ \\
    \hline
    \hline
\end{tabular}
\caption{Results of the comparison between the significance obtained for TXS 0506+056 when using an analysis with features similar to the one in \cite{IceCube:2018cha} and the one presented in this paper. When testing the impact of different datasets, the years 2012-2015 are analyzed with the multi-flare algorithm. 
When testing the impact of a different strategy in the combination of the samples, the single-flare algorithm is used on the dataset \MA{{\tt PSTracks v3}}: in one case only the IceCube sample containing the known flare is analyzed and the p-value penalized, adopting the same logic as in~\cite{IceCube:2018cha}; in the other case all the 10-year samples are combined in a joint likelihood, as described in Section~\ref{sec:analysis}, and no penalization is needed.}
\label{tab:TXS_comparisons}
\end{table}


\section{Discussion on the multi-messenger context}
\label{sec:MM}

As shown in Section~\ref{sec:results}, M87 is the most significant source of the catalog, exhibiting 3 events over a time lag of minute scale with post-trail significance of $1.7~\sigma$. It is one of the closest ($z=0.00436$) potential cosmic ray accelerators, hosting a supermassive black hole of $6.5\times10^9M_\odot$. Its jet was already observed more than a century ago ~\citep{blanford_agn} in a large elliptical radio galaxy of Fanaroff-Riley type I in the Virgo cluster.
It has been observed in $>100$~GeV energy region: VERITAS detected a flare extending beyond 350~GeV with a spectral index at the peak of $2.19 \pm 0.07$ \citep{Aliu_2012} in Apr. 2010. In a 2008 flare, a clear correlation between the X-ray emission and the TeV one \cite{Acciari_2008,Albert:2008kb}. Previous positive detection was reported by HEGRA in 1998/99 above 700 GeV~\citep{2003A&A...403L...1A} , and up to $\sim 10$~TeV by H.E.S.S. in 89 hours of observation between 2003-6, showing a variability at the time scale of a few days in the 2005 high state associated to the Schwarzschild radius of M87 \cite{Aharonian_2006}. Recently, MAGIC reported the results on the monitoring of M87 for 156 hours in 2012-15 \cite{MAGIC2020}. It is worth noting that HAWC set an upper limit above 2 TeV for 760 d of data. The non-observation of gamma-rays at $>$~TeV energies, may indicate a cut-off in the spectrum. Such cut-off may differ for neutrinos, being less affected by the absorption in the source and by the extra-galactic background light. 

\begin{figure}[htbp]
	\centering
	\includegraphics[width=0.9\linewidth]{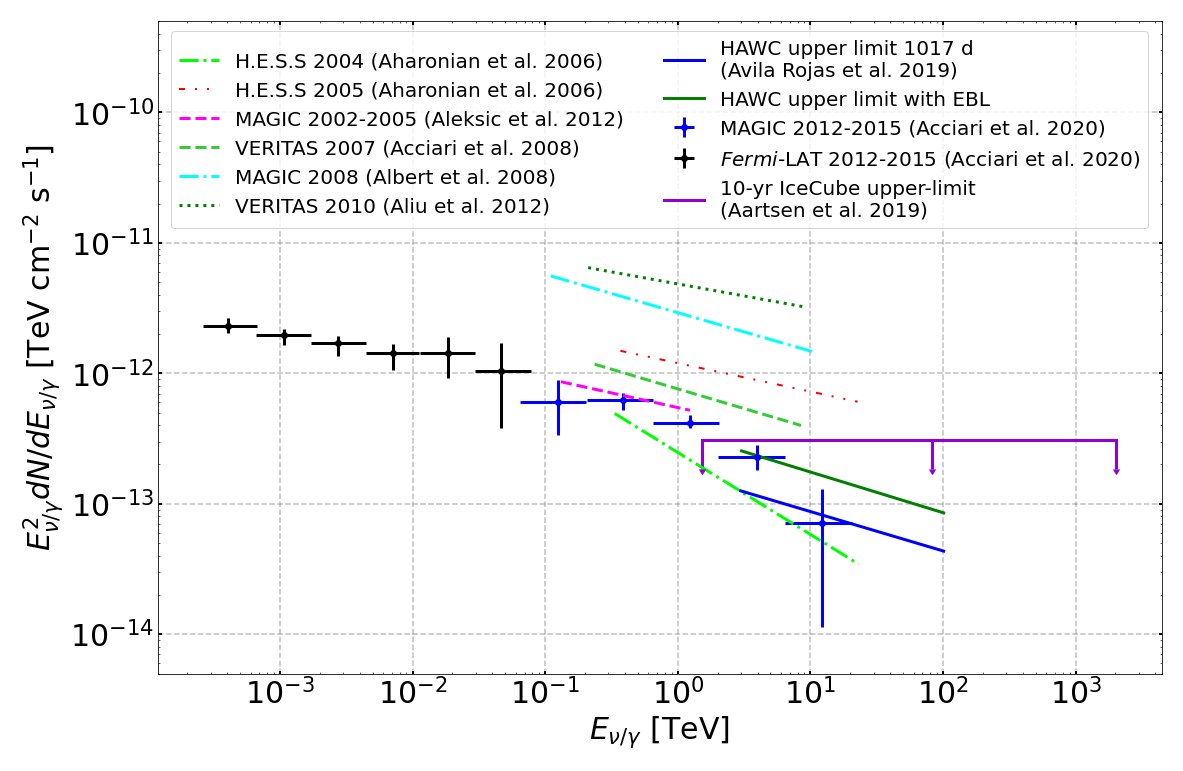} 
	\caption{The gamma-ray flux in the steady state of the source observed between 2012-2015 \cite{MAGIC2020} is shown with black (Fermi-LAT) and blue dots (MAGIC). The higher level dashed lines are levels of flux observed during flares (see references in the text). The purple line with downing arrows corresponds to the 10-year time-integrated upper limits taken from ~\cite{Aartsen:2019fau}, with an assumed spectrum $dN/dE\sim E^{-2}$.}
\label{fig:MM}
\end{figure}

The gamma-ray observations from M87 are summarized in Fig.~\ref{fig:MM}, together with the 10-year time-integrated upper limits on the neutrino flux estimated in~\cite{Aartsen:2019fau} for a spectrum of the form $dN/dE\sim E^{-2}$. 

Precise radio observations \cite{Sikora_2016} indicate a persistent central ridge structure, namely a spine flow in the interior of M87 jet, in addition to the well-known limb-brightening profile, which needs further measurements. A composite structure of the jet has been speculated also for TXS 0506+056 based on observations months after the detection of the IceCube high-energy event that triggered its multi-wavelength observations. With the millimeter-VLBI it was observed that the core jet expands in size with apparent super-luminal velocity \cite{Ros:2019bgo}. This can be interpreted as deceleration due to proton loading from jet-star interactions in the inner host galaxy and/or spine-sheath structure of the jet \cite{2005A&A...432..401G,Tavecchio:2008be}.  This sort of spine-sheat structure has been advocated as a possible explanation for the higher flux of neutrinos than gamma-rays and also suggested for TXS 0506+056 by MAGIC \citep{2018ApJ...863L..10A}, while models with a single zone struggle to explain the 2014-2015 flare of TXS 0506+056 (see e.g. \cite{Murase_2018,Zhang:2019htg,2018ApJ...864...84K}).

Other models, e.g \cite{Inoue:2019yfs,Murase_2020}, have been revised to explain the more recent observations of IceCube on NGC 1068 \citep{Aartsen:2019fau}. These models focus on the higher observed flux of IceCube neutrino events in the $\sim 1-50$~TeV region with respect to the level of gamma-ray fluxes observed at lower energy by Fermi and the limits of MAGIC. The corona super-hot plasma around the super-massive black hole accelerates protons, carrying few percent of the thermal energy, through plasma turbulence \cite{Murase_2020} or shock acceleration \cite{Inoue:2019yfs} leading to the creation of neutrinos and gamma rays. The environment is dense enough to prevent the escape of $\gg$ 100 MeV gamma rays while $\sim \mathrm{MeV}$ gamma-rays would be their result from cascading down.
Further insights will be needed in both messengers and all wavelengths to better constrain the structure of jets and acceleration mechanisms in one or multiple zones.

\newpage
\bibliography{main}{}
\bibliographystyle{aasjournal}



\end{document}